\documentclass{article}
\usepackage[round,authoryear]{natbib}
\usepackage{arxiv}
\usepackage{graphicx} 
\usepackage{url}
\usepackage{subcaption} 
\usepackage[utf8]{inputenc} 
\usepackage[T1]{fontenc}    
\usepackage{hyperref}       
\usepackage{url}            
\usepackage{booktabs}      
\usepackage{svg}            
\usepackage{csquotes}
\usepackage{todonotes}
\usepackage{tabularx}
\usepackage{verbatim}
\usepackage{enumitem}
\usepackage{amsmath} 
\usepackage{longtable}
\usepackage{booktabs}

\usepackage[edges]{forest}
\usetikzlibrary{shadows.blur}
\usetikzlibrary{positioning,arrows.meta}
\usetikzlibrary{trees}

\newcommand{\detailtexcount}[1]{%
  \immediate\write18{texcount -merge -sum -q #1.tex > #1.wcdetail }%
  \verbatiminput{#1.wcdetail}%
}

\newcommand{\detailtexcountsection}[1]{%
  \immediate\write18{texcount -merge -sub=section -q #1.tex > #1.wcdetail }%
  \verbatiminput{#1.wcdetail}%
}

\title{Keeping an Eye on AI: A Framework for Effective Human Oversight of AI Systems}
\renewcommand{\shorttitle}{Keeping an Eye on AI: A Framework for Effective Human Oversight of AI Systems}

\author{
  \textbf{Susanne Gaube}\thanks{Corresponding author: susanne.gaube@ucl.ac.uk} \\
  University College London \\
  \And
  \textbf{Markus Langer}\thanks{Corresponding author: markus.langer@psychologie.uni-freiburg.de} \\
  University of Freiburg \\
  \And
  \textbf{Tim Miller}\thanks{Corresponding author: timothy.miller@uq.edu.au} \\
  University of Queensland \\
  \AND 
  \normalfont
  Kevin Baum$^1$, Raimund Dachselt$^2$, Anna Maria Feit$^1$, Ujwal Gadiraju$^3$, Harmanpreet Kaur$^4$,\\
  Mark T. Keane$^5$, Richard Landers$^4$, Johann Laux$^6$, Q. Vera Liao$^7$, Brian Lim$^8$, \\
  Linda Onnasch$^9$, Tim Schrills$^{10}$, Liz Sonenberg$^{11}$, Chenhao Tan$^{12}$, Nava Tintarev$^{13}$, \\
  Ziang Xiao$^{14}$, Hanwei Zhang$^1$
  \vspace{0.2cm} \\
  \small $^1$Saarland University, $^2$TU Dresden, $^3$Delft University of Technology, $^4$University of Minnesota, \\
  \small $^5$University College Dublin, $^6$University of Oxford, $^7$University of Michigan, $^8$National University of Singapore, \\
  \small $^9$Technische Universität Berlin, $^{10}$University of Lübeck, $^{11}$University of Melbourne, \\
  \small $^{12}$University of Chicago, $^{13}$Maastricht University
}

\date{March 2026}

\begin{document}
\maketitle

\begin{abstract}
    The use of Artificial Intelligence (AI) in high-risk, decision-making scenarios presents technical, safety, and normative challenges; problems that may only be ameliorated by human oversight. However, notions of human oversight lack a common foundational understanding: oversight architectures are not well defined, the roles involved remain unclear, and implementation steps are opaque. Hence, researchers and practitioners struggle to determine how to design, implement, and evaluate systems that enable effective human oversight. This paper advances a practical framework for effective human oversight of AI systems, based on a cross-disciplinary perspective that draws on insights from computer science, human-computer interaction, psychology,  philosophy, and law.  The core contributions are: (1) a foundational framework, with a working definition, architecture and processes for effective human oversight of AI systems; (2) an initial template for documenting oversight architectures and processes, applied to diverse domains; and (3) a synthesis of open research challenges that need to be considered in the emerging field of effective human oversight of AI systems.

\end{abstract}

\section{Introduction}



Artificial Intelligence (AI) is increasingly being deployed across domains, impacting industry, society, and private lives. AI-enabled technologies can and will deliver many positive impacts, but their benefits also come with risks ~\citep{cummings2024taxonomy,olteanu2025rigor,shelby2023sociotechnical}.
Harmful outcomes can occur in various ways. For example, model drift can degrade performance over time, technical problems can cause failures in unexpected situations or complete operational breakdown (e.g., during power outages) \citep{vela2022temporal, javed2024robustness, lasko2024probabilistic}). AI systems also face difficult-to-detect security threats such as adversarial attacks~\citep{tully2025patient, brohi2025ai}. Beyond technical concerns, normative risks emerge when system behaviour leads to unfair, discriminatory, degrading, or privacy-violating outcomes \citep{celi2022sources, alderman2025tackling, murdoch2021privacy, karimian2022ethical, yang2024limits}. In high-risk domains such as healthcare or critical infrastructure, such risks can have life-threatening consequences. Human oversight is often proposed as a safeguard against these risks, but what such oversight should entail is typically unspecified.

This paper aims to make the notion of human oversight concrete by proposing a framework for human oversight that mitigates the risks from AI-enabled technologies in high-risk use cases. In part, our work is a response to regulations that explicitly require human oversight of AI; for instance, Article 14 of the EU's AI Act mandates that high-risk AI systems must be overseen by natural persons, as a way to mitigate residual risks that technical safeguards cannot eliminate \citep{cheong2024transparency,McBride02112014,dearteaga2020acaseforhumans}. In line with prior research, human oversight comprises monitoring AI systems, detecting inaccurate or inadequate outputs, recognizing failures, spotting malicious manipulation, making normative judgments about justice and fairness, and intervening when deemed necessary \citep{langer2024effective,Langer2026,van2024challenges,zerilli2019algorithmic,green2022flaws,sterz2024quest,enqvist2023human,laux2024institutionalised}. Yet the significance of human oversight extends beyond regulatory compliance and jurisdictional boundaries. Across diverse domains, it is increasingly regarded as a critical---and perhaps final---safeguard against the manifold risks posed by AI systems, and as a critical pillar in the safety and accountability infrastructure for responsible AI.

While the conditions for effective human oversight of AI are yet to be understood, it is evident that oversight can fail easily. Human overseers can fail to detect inaccurate outputs or fail to intervene effectively \citep{rieger2025highly, vaccaro2024combinations, gaube2021ai}. Even more concerning, human oversight itself can introduce inaccuracies and biases \citep{rieger2025highly,green2019disparate}, especially when control mechanisms are insufficient \citep{sterz2024quest}. In other scenarios, oversight may not be humanly possible \cite{green2022flaws}. Most critically, poorly-designed oversight can create the illusion of human control, when oversight structures give humans the responsibility, but not the agency, to prevent harm \citep{Romeo_2025, sterz2024quest}. 

Given these challenges, many fundamental questions about human oversight need to be answered, including: What should the goals of human oversight be? What oversight activities need to be carried out? Who is qualified to perform this oversight? Under what conditions can it be effective and meaningful? And what happens when effective human oversight is impossible or infeasible? Some of these questions are not new; decades of research on human–automation interaction \citep{mosier2019humans,sheridan2021chapter, Bainbridge_1983} have addressed similar issues in aviation, nuclear power, industrial production, and autonomous weapons systems \citep{McBride02112014,schwarz2021autonomous} (where a recent NATO report shows the depth of the discussions on human oversight and control \cite{draper2020human}). However, the scale, complexity, and opacity of today's AI systems---including foundation models and agentic systems that act autonomously across multiple steps---present new challenges, requiring new forms of oversight based on interdisciplinary understandings ~\citep{cummings2023frontiers,endsley2023ironies,naikar2023, narayanan_2026designing}. Prominent challenges for oversight of AI systems include \citep{langer2024effective,pritchett2024things}:  increased agency that can lead to new anomalies and, hence, new risks \citep{Acharya_2025}; the probabilistic, opaque nature of AI systems results in new level of complexity, including inter-dependencies (e.g., in multi-agent systems \citep{hammond2025multiagentrisksadvancedai});  and changes in model effectiveness over time (e.g., model drift \citep{van_der_Vorst_2025}. Furthermore, such challenges occur against the backdrop of shorter development-to-deployment cycles \citep{Bengio_2024} across a wide variety of emerging application domains. 
These factors make it difficult to identify the capabilities required of operators, overseers, and organisations to build effective oversight architectures and processes, and motivate the definition of a principled, general framework for human oversight of AI. Such a framework must build on prior research while acknowledging that the scope, complexity, and societal stakes of human oversight have fundamentally changed.

This paper integrates insights from computer science, human-computer interaction, philosophy, ethics, psychology, human factors, and law to develop a common ground for the emerging research field of human oversight of AI. It is the result of cross-disciplinary discussions at the 2025 Dagstuhl Seminar on Human Oversight of AI Systems (see \cite{langer_challenges_2026}). Achieving effective human oversight requires well-designed oversight architectures, tools, and processes, as well as means to evaluate their effectiveness. To this end, our core contributions are:

\begin{itemize}
\item[(1)]  Development of a working definition (Section \ref{sec:foundations}) and foundational framework (Section \ref{sec:framework}) for effective human oversight of AI.
\item[(2)] Operationalisation of this framework in a template, demonstrably applied to diverse domains (Section \ref{sec:template}).
\item[(3)] Identification of open challenges to be addressed in future research on human oversight of AI (Section \ref{sec:futurework}).
\end{itemize}

\section{Grounding Human Oversight}
\label{sec:foundations}




Human oversight can be viewed as a control and safety layer, ideally one that prevents unfavourable outcomes from AI systems, and critically embedded in a larger risk management architecture \cite{rasmussen1997risk} given that human oversight cannot address all risks. From this perspective, \textbf{Human Oversight of an AI system} can be defined as a:

\begin{quote}
\textbf{Deliberate} human activity with the goal of sufficiently mitigating risks in the use of an AI system for a specific task. This oversight, which may be technologically supported,  often works within \textbf{distributed} organisational structures and processes that have to meet regulatory and economic requirements. It requires one or more people to \textbf{monitor} the operation and output of the AI system and to \textbf{intervene} when professional judgment, established criteria, or legal provisions indicate that action is required to mitigate risks arising from the system's processes, outputs, or its context of use. Its \textbf{effectiveness} is assessed by the reduced  likelihood of occurrence and severity of harms arising from the system operations.
\end{quote}

This definition applies to decisions that are 
dynamic and time-critical (e.g., overseeing autonomous vehicles) and high-stakes, deliberative (e.g., medical, judicial, safety \citealp{cummings2024taxonomy, olteanu2025rigor, shelby2023sociotechnical}). 
Below, we briefly expand on some key considerations in this definition:

\begin{description}
\item [Deliberate]
Human oversight does not arise incidentally. It requires explicit assignment: individuals must be designated to oversight roles, trained accordingly, and embedded within organisational structures. 

\item [Distributed] Human oversight may also be distributed across individuals, hierarchical layers, and organisations, which may often be a key requirement for effective oversight \citep{Grote_2024}. An operator may monitor single outputs, a domain expert may review borderline cases, developers may analyse AI system behaviour over time, and a compliance officer may monitor communication between the frontline operators and developers. Tasks with clear goals and responsibilities need to be explicitly allocated to these parties. 

\item [Monitor]
Monitoring is the ongoing process of observing (potentially aggregated) system behaviour and outputs against relevant benchmarks,
such as pre-defined performance thresholds, legal standards, or professional norms.

\item [Intervene] Intervention is the action taken when monitoring reveals risks: correcting an output, halting a process, flagging a case, escalating a decision, or re-training or re-designing a system. Crucially, the capacity to intervene is predicated on the overseer having both the understanding to recognize when action is needed and the control or causal influence to execute that action \citep{sterz2024quest}. 

\item [Effective]
Effective human oversight \enquote{reduces occurrence and severity of harms}. This can be realised in three increasingly demanding ways: \textit{handover}, \textit{error detection}, and \textit{synergy}. \textit{Handover} involves humans taking over when the AI system fails, or reaches the boundary of its competence. This required sufficient situational awareness to ensure safe transfer of control \citep{endsley2023ironies}. \textit{Error Detection} involves  identifying inadequate system behaviour and deciding whether to override the AI or not \citep{langer2024effective, rieger2025highly}. Its effectiveness depends on the detectability of errors, overseer capability to identify them, and on the efficacy of available interventions. 
\textit{Synergy} requires \textit{appropriate reliance}---intervening on erroneous outputs while trusting well-functioning AI---, up to the point where human-AI performance exceeds either alone\citep{vaccaro2024combinations, gaube2026underreliance, eckhardt_2025asurvey}). Here, oversight reduces both AI-related risks and risks introduced by human intervention.

\end{description}

\textbf{Enabling human oversight effectiveness.}
Establishing, ensuring, and evaluating the effectiveness of human oversight is a socio-technical challenge, requiring the integrative design of human, technological, and organisational factors (as argued by e.g. \citealp{cheong2024transparency,Kaber2025,naikar2023,rieger2025highly,sterz2024quest,laux2024institutionalised}). Our framework therefore, adopts a layered approach that takes into account the following factors:

\begin{itemize}[leftmargin=*]

\item\textit{Human factors} capture the knowledge, skills, motivation, and behaviours of oversight personnel. 
These need to be aligned with the oversight task. Oversight personnel includes individuals trained in taking over control of AI, experts evaluating single AI outputs and edge cases, or people with expertise on technical issues and failures. For example, system developers might monitor technical functions using benchmark information, whereas domain experts might evaluate outputs using domain knowledge and context-specific insights. 

\item\textit{Technical factors} cover improving model reliability, failure or risk detection, and system interfaces for overseers, including considering human-interaction modes, effective visualisations, and warning modules supporting situational awareness.
In particular, providing transparency into the AI system, including metrics related to the risks being monitored, explainability, and uncertainty communication may be prerequisites for effective oversight.

\item\textit{Organisational factors} relate to the work environment, the available time and resources, and establishing 
clear responsibilities for and between actors. 
For example, oversight organisation requires defining the concrete tasks of oversight personnel, whether oversight personnel works individually or in teams; in a single organisation or across organisations. 

\end{itemize}

\section{A Framework for Human Oversight of AI Systems}
\label{sec:framework}



We propose a framework that casts human oversight as a deliberate, evidence-informed layer of the safety and control infrastructure of AI systems. Thus, human oversight is realised within a complex socio-technical system with layers involving staff, management, organisations, regulators/associations, and government (echoing Rasmussen’s six-layer Risk Management Framework \citep{rasmussen1997risk} for accident analysis). 

Our framework makes a critical distinction between human oversight architecture and process. The \textit{Oversight Architecture} (Section \ref{sec:architecture}) is a static model that identifies the key components of the framework. The \textit{Oversight Process} (Section \ref{sec:process}) is a dynamic model that identifies the flow of information between different components, notably those involving \textit{monitoring} and \textit{intervention} (Section \ref{sec:monit_interv}). 

There are two main layers in the framework, the \textit{task layer}, which covers the task realised with the AI system, and the \textit{oversight layer}, that covers the human oversight of the AI-enabled task layer. Figure~\ref{fig:HO_two_layer_architecture} shows these two layers with information flows between them for monitoring and intervention; more precisely, the task layer is embedded within the oversight layer (see Figure~\ref{fig:HO_process}). Importantly, as we will discuss in Section \ref{sec:unpacking}, the oversight layer may be modeled as further layers of human oversight to become effective (cf. Figure~\ref{fig:HO_two_layer_to_multi_layer_architecture}).

\subsection{Human Oversight: The Architecture} \label{sec:architecture}

\begin{figure}[!th]
  \centering
  \includegraphics[width=0.70\textwidth]{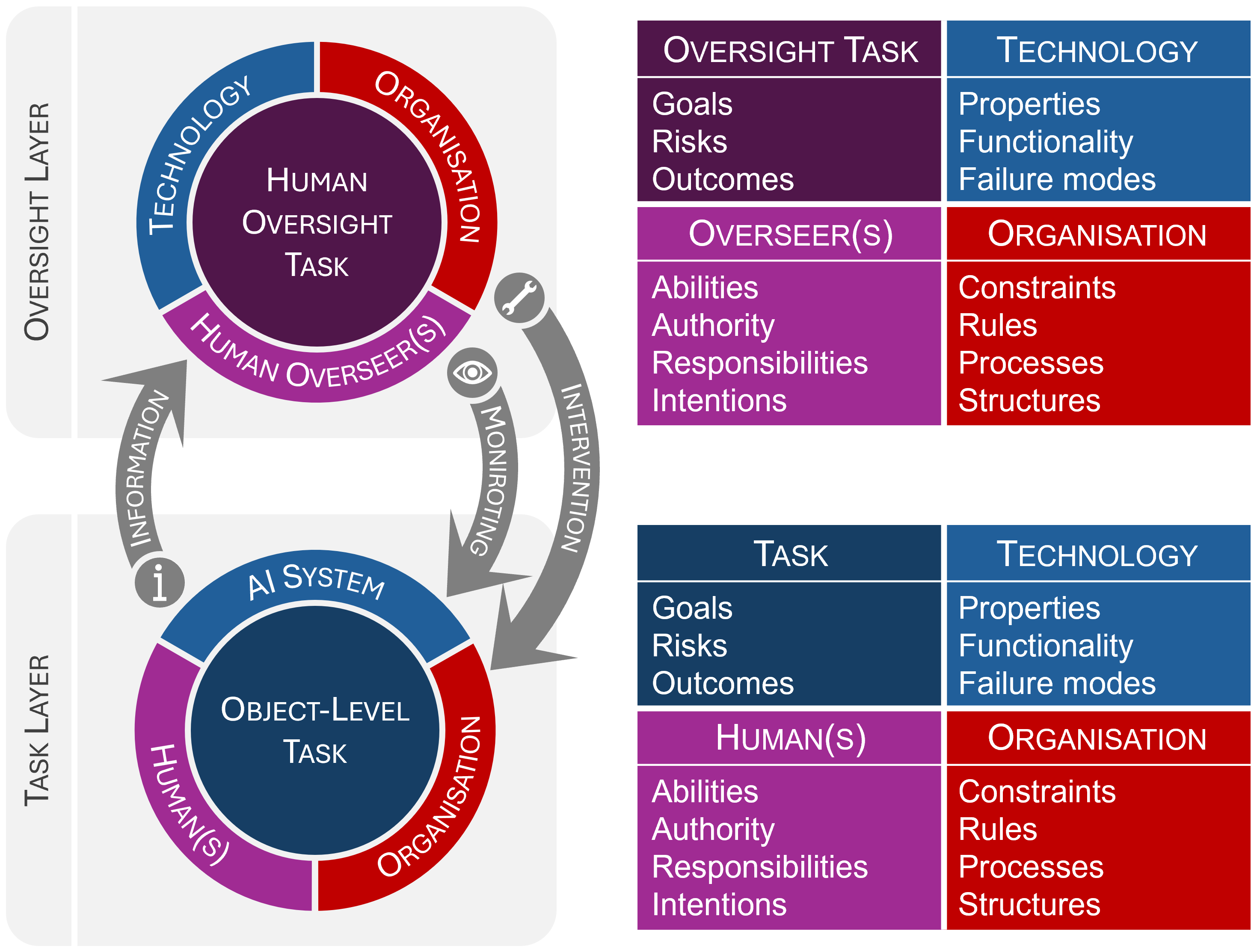}
  \caption{An architecture for human oversight. The two components of the architecture are the \textit{oversight layer} and the \textit{task layer}. The task layer (lower-left) reflects the socio-technical system that is to be overseen, involving some AI system executing or assisting a task with human users/operators (optional), and the organisation of the task. The oversight layer (upper-left) reflects the socio-technical system performing the oversight with a human overseer(s) who may be supported by technology (optional) to achieve oversight goals within an organisational setting. To the right we see boxes with illustrative examples of parts of both of these layers. 
  }
  \label{fig:HO_two_layer_architecture}
\end{figure}

Figure~\ref{fig:HO_two_layer_architecture} shows the main idea of the architecture in which the \emph{oversight layer} is there to monitor the \emph{task layer}, and may intervene on the task layer. Both layers capture the socio-technical nature of the respective task, integrating the technical, human, and organisational factors that impact effective human oversight.

\subsubsection{The Task Layer}
The task layer contains the object-level task being executed or assisted by the AI system, along with its wider socio-technical aspects (see Figure~\ref{fig:HO_two_layer_architecture}, lower-left):

\begin{itemize}[leftmargin=*]
    \item \textit{Object-Level Task}: The task being done/assisted by the AI system; its goals, outcomes occurring, and  risks arising.
    \item \textit{AI System}: The AI system doing the task; its properties, functionality and failure modes.
    \item \textit{Human(s)} (optional): The human(s) doing the task with the AI system; their abilities, authority, responsibilities and intentions.
    \item \textit{Organisation}: The constraints, rules, processes, and structures in which the task is undertaken---not necessarily a literal organisational entity, but the organisational factors affecting the task.
\end{itemize}

Table~\ref{tab:architecture_example} shows how this task layer can be instantiated in a sample diagnostic decision-making scenario.

\begin{table}[!ht]
\centering
\caption{Simplified Diagnostic Decision-Making Scenario: Task \& Oversight Layers}
\label{tab:architecture_example}
\setlength{\tabcolsep}{3pt} 
\renewcommand{\arraystretch}{1.2}
\begin{tabular}{p{0.2\textwidth}p{0.75\textwidth}}
\toprule
\textbf{Component} & \textbf{Example}\\
\midrule
\multicolumn{2}{c}{\textbf{\textit{Task Layer}}}\\

\textit{Object-Level Task}
& The task goal is to detect an acute haemorrhage in a head CT scan and provide a diagnosis, where potential risks include missed findings, delayed diagnosis, and unnecessary medical interventions\\

\textit{AI System}
& The AI tool that analyses CT scans, highlighting suspect regions, and providing haemorrhage-probability scores\\

\textit{Human(s)}
&  The users of the AI tool, the radiologists or emergency clinicians who review AI flags, interpret images, and make time-critical decisions on medical intervention (n.b., fully-automated triage there could be minimal human involvement) \\

\textit{Organisation} 
& The associated clinical protocols, IT systems (e.g., PACS -- Picture Archiving and Communication System), and governance for the AI deployment\\

& \\
 \midrule
\multicolumn{2}{c}{\textbf{\textit{Oversight Layer}}}\\

\textit{Human Oversight Task}

 &  To mitigate risks such as rectifying diagnostic errors from AI or fixing workflow issues in CT scan interpretation \\

\textit{\mbox{Technology}}
& The oversight software that records diagnoses for audits or flags the occurrence of potential misdiagnoses\\

\textit{\mbox{Personnel} }
& The senior radiologist auditing decisions or detecting issues with diagnostic outputs and clinical safety officers monitoring AI drift\\

\textit{\mbox{Organisation}} 
& Available time allocated to review AI outputs; communication channels between frontline radiologists and senior radiologists; distributed responsibilities between the senior radiologist and the clinical safety officer;  \\

\bottomrule
\end{tabular}
\end{table}

\subsubsection{The Oversight Layer}
The oversight layer contains the task of overseeing, along with its wider socio-technical aspects (see Figure~\ref{fig:HO_two_layer_architecture}, upper-left):

\begin{itemize}[leftmargin=*]
    \item \textit{Oversight Task}: This task contains goals to mitigate risks to safety, health, and fundamental rights and the outcomes of oversight (e.g., effective risk mitigation); optionally, there can be multiple oversight goals (e.g., safety as a primary goal and operational efficiency as a secondary goal).
    \item \textit{Oversight Technology} (optional): The technology supporting the human oversight task; for instance, an automated warning system to alert overseers when key measures are out of bounds.
    \item \textit{Oversight Personnel}: Those humans assigned oversight duties, such as domain experts in the object-task domain, or people specifically trained to monitor AI systems.
    \item \textit{Oversight Organisation}: The organisational constraints, rules, processes, and structures that the human overseers need to consider; for instance, oversight personnel could work as a team in a control centre, specifically designed for oversight duties, where each person has assigned responsibilities.
\end{itemize}

From a process perceptive, here human overseers are \textit{monitoring} information flows from the task layer; for example, they may use system logs to check system performance. They are also \textit{intervening} in the task layer; for example, to overwrite inaccurate system outputs or call for technical support to update the system (see also Figure~\ref{fig:HO_process}) showing how the task layer is actually embedded within the oversight layer. Table~\ref{tab:architecture_example} shows how this oversight layer might be instantiated in a sample Diagnostic Radiology scenario.

\subsubsection{Unpacking the Oversight Layer}
\label{sec:unpacking}

\begin{figure}[!ht]
  \centering

  \includegraphics[width=0.75\textwidth]{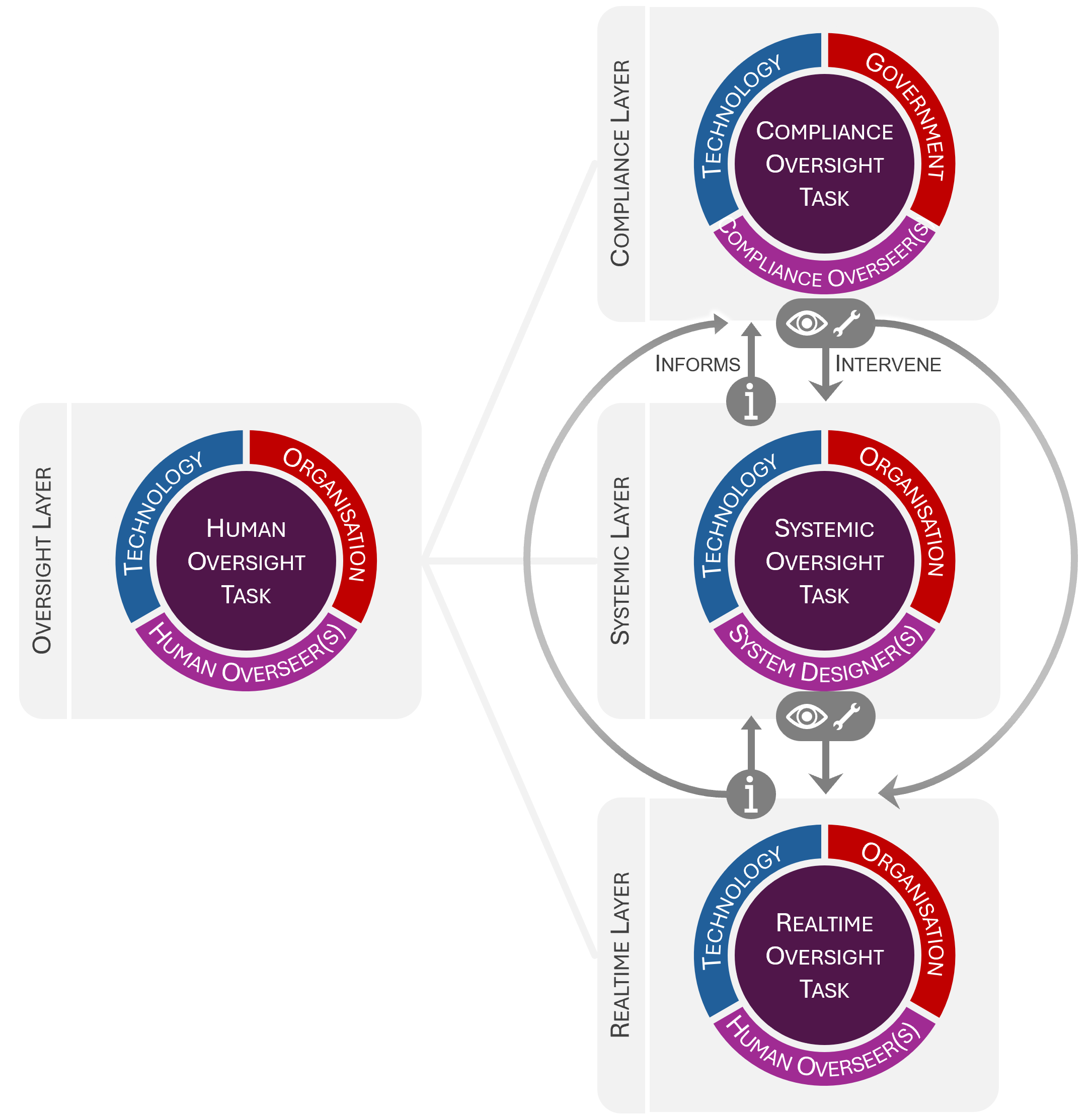}
  \caption{An Oversight Layer may need to be unpacked into different layers of oversight; there may be a \textit{Real-Time Layer} (i.e., with overseers monitoring/intervening on outputs in real-time), a \textit{Systemic Layer} (i.e., with developers monitoring/intervening, e.g. by overseeing AI functioning on an aggregate level), and a \textit{Compliance Layer} (i.e., with compliance officers monitoring/intervening to enforce regulations/contracts). Note, these different layers of oversight will be affected by other even broader layers of risk-mitigation in AI deployment; e.g., legal provisions.}
  \label{fig:HO_two_layer_to_multi_layer_architecture}
\end{figure}

 The above presentation of the oversight layer may be a simplified view. In order to ensure effective human oversight, often this layer needs to be unpacked into several layers of oversight (see Figure~\ref{fig:HO_two_layer_to_multi_layer_architecture}). We argue for, at least, three such layers:

\begin{itemize}[leftmargin=*]
    \item \textit{Real-time Oversight:} Continuous monitoring of the task layer in real-time, with dynamic control interventions to halt, modify, or override single AI outputs before harm or non-compliance occurs. For instance, real-time oversight could be a radiologist correcting errors in an AI-generated report for a specific patient.
    \item \textit{Systemic Oversight:} Monitoring the object-level task to perform interventions that recalibrate, redesign, or retrain the AI system; or interventions that modify the attributes of the task, human(s) or organisation, to be better aligned with the oversight goals, to prevent future harm or non-compliance. For instance, systemic oversight could be a system designer monitoring for AI model drift over multiple tasks to eventually retrain a system.
    \item \textit{Compliance Oversight:} Monitoring the object-level task in real-time to perform interventions that enforce regulatory and contractual obligations. For instance, compliance oversight could be a compliance officer in a hospital using patient reports to mandate model re-training.
\end{itemize}

These considerations show that oversight is not just restricted to real-time interventions, but can also occur over longer time-periods with monitoring and intervention activities that may require access to information or capabilities that are available in the systemic or compliance layers. For instance, longer-term oversight often acts on aggregate data---over days or weeks---to prevent \emph{future} harm or non-compliance (e.g., monitoring group fairness in AI decisions across many tasks, rather than individual decisions \citep{langer2024effective}).  

We acknowledge that other `layers' beyond these can affect \textit{human} oversight of AI systems by being e.g.  part of a larger risk-mitigation ecosystem. For example, legal aspects can impact the design, organisational policy, and normative judgments made by oversight personnel \citep{rasmussen1997risk}. These different layers and the broader risk-mitigation layer highlight the complex inter-dependencies that affect human oversight.

\subsection{Human Oversight: The Process }
\label{sec:process}

\begin{figure}[!th]
  \centering
  \includegraphics[width=\textwidth]{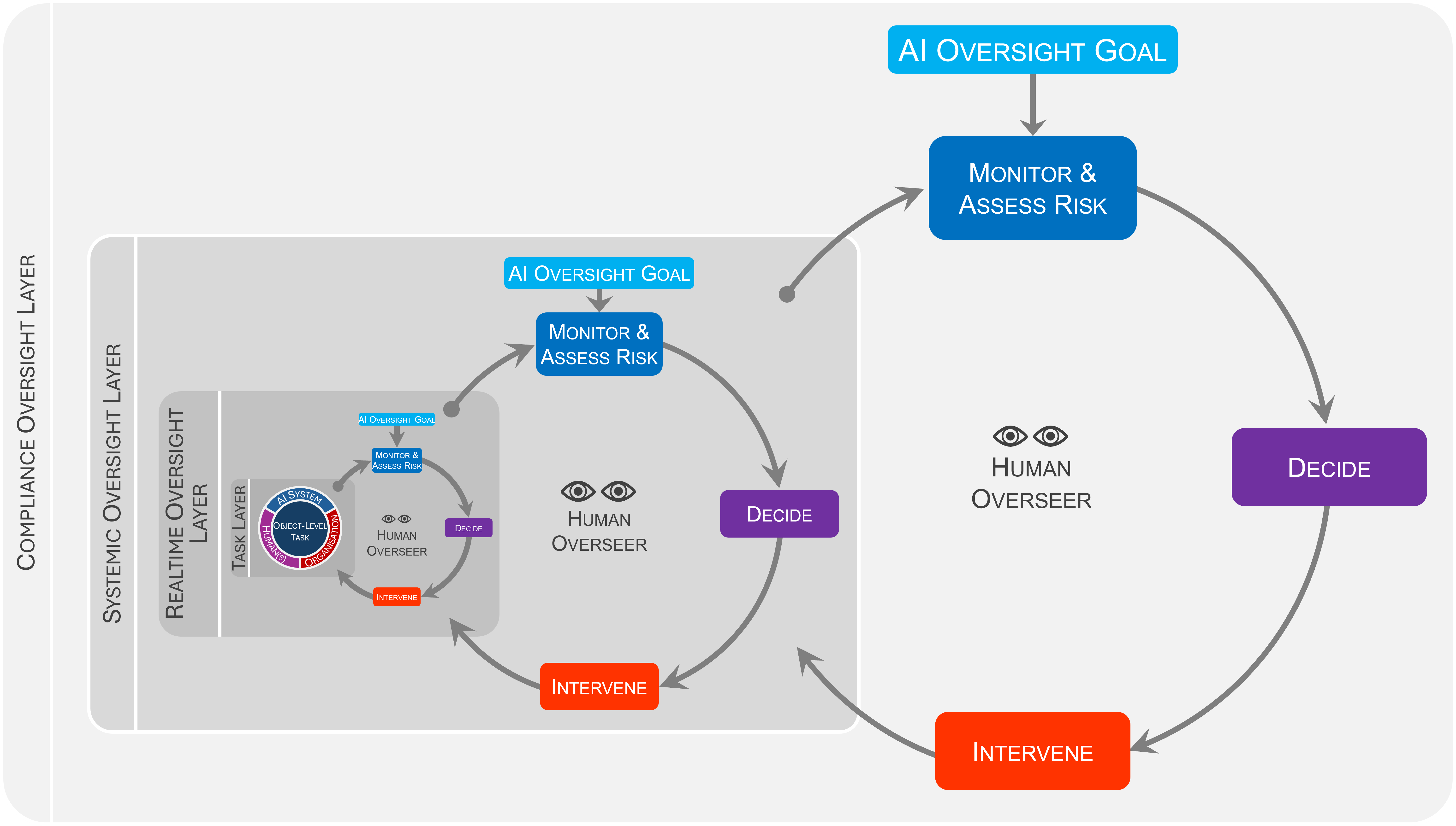}
\caption{Human Oversight Process. Oversight layer with an \emph{observe-think-act process} of monitoring, deciding, and intervention. While specific responsibilities change between layers in specific contexts, the general process is the same in all layers.}
  \label{fig:HO_process}
\end{figure}

Figure~\ref{fig:HO_process} shows the \textit{Human Oversight Process} of our framework. It is modelled as a cybernetic control loop, that is, a self-regulating system in which actions are continuously adjusted on the basis of feedback about their effects to maintain or move the system towards some desired state \citep{carver2000structure}. In this loop, overseers \emph{monitor} the task layer and gather information to \emph{assess risks} against the \emph{goals} of oversight (e.g., a safety risk), and then \emph{decide} on \emph{interventions} (e.g., they may override outputs and halt the system or they may do nothing). Table~\ref{tab:human_oversight} shows an instantiation of an oversight process in a sample Diagnostic Radiology scenario.

\begin{table}[htbp]
\centering
\caption{Human oversight processes, their descriptions, and design options.}
\label{tab:human_oversight}
\setlength{\tabcolsep}{4pt} 
\renewcommand{\arraystretch}{1.2}
\begin{tabular}{p{0.1\textwidth}p{0.35\textwidth}p{0.5\textwidth}}
\toprule
\textbf{Process} & \textbf{Design Options for Human Oversight}& \textbf{Example} \\
\midrule

\emph{Monitor} &  Support detection and interpretation of inaccurate and inadequate outputs; support data aggregation and visualisation& Radiologist monitors AI outputs and detects false positives/negatives; performance dashboards summarise trends for Clinical Safety Officers\\

\emph{Assess Risk} & Tools to simulate consequences of inaccurate AI support for a larger scale of cases &  Radiologist decides that inaccurate output is a patient safety risk; Clinical Safety Officer reviews emerging patterns, considers the clinical and operational impact of false positives/negatives (e.g., over/under treatment)\\

\emph{Decide} & Offer clear decision guidelines for whether and when to intervene& Radiologist decides to override the AI output; Radiology Governance Committee limits AI use to certain patient groups  \\

\emph{Intervene}  & Enable effective communication across oversight layers & Radiologist changes report generated by the AI tool; Clinical Lead suspends AI use for affected indications, informs Digital Health team, and reports to the vendor\\

\bottomrule
\end{tabular}
\end{table}

Interventions will affect the task layer that may, in turn, lead overseers to gather new information in ongoing monitoring. This nested loop shows the task layer embedded within the oversight layer and how the two can interact. Interventions may also happen between layers; for example, an overseer in a real-time layer (e.g., operator of the system) could raise a flag to the systemic layer requesting oversight and intervention from that layer (e.g., from a developer) when they detect risks that they cannot mitigate themselves. Likewise, they could pass information to another layer to inform this other layer's risk assessment (see the intervene-arrow going outside the layer to another layer in Figure~\ref{fig:HO_process}).

Furthermore, in this process model, the human-entities represent roles not individuals; that is, the same person could fill more than one role (e.g., a `user' and `overseer'). For example, a radiologist using a diagnostic system could detect inaccurate AI outputs and then intervene to override specific decisions while also reporting these anomalous patterns to other oversight layers (e.g., the systemic layer). So, this person has a role in the task layer (the inner loop of Figure \ref{fig:HO_process}) and in the real-time oversight layer (the outer loop of Figure \ref{fig:HO_process}).

\subsection{Monitoring \& Intervention Processes}
\label{sec:monit_interv}

The two fundamental processes underlying human oversight are monitoring and intervention \citep{langer2024effective, McBride02112014, sheridan2021chapter, Sheridan_1978}. Their successful implementation is essential for meeting the requirements of effective human oversight \cite{sterz2024quest}. Specifically, monitoring is related to the epistemic requirements of effective human oversight, encompassing the oversight personnel's awareness of the operational environment, system states, processes, outputs, associated risks, available interventions, as well as their respective benefits and likely consequences. Intervention, in turn, addresses the requirements concerning human causal influence and control over the system including the opportunity, capabilities, and resources necessary to stop, override, or modify AI processes and outputs. Below, we introduce high-level operational preconditions for successful monitoring and intervention processes

\paragraph{Monitoring}
Monitoring AI systems, assessing current risks, and detecting problematic system behaviour can be understood as a \emph{signal detection task} \citep{langer2024effective, McBride02112014}. Against the backdrop of theories of human perception and judgment such as Signal Detection Theory and Brunswik’s Lens Model \citep{green1966signal, brunswik1956perception, kellen2018elementary}, conceptualising monitoring as signal detection highlights that oversight personnel search for information cues in the task layer that provide evidence for risk or problematic behavior. During monitoring, oversight personnel thus accumulate evidence when they attend to, detect, and interpret cues until the available evidence warrants intervention. Whether monitoring is successful depends on the availability and quality of information cues, as well as the ability of the oversight to interpret these \citep{SCHLICKER2025108671, funder1995accuracy}.

From this perspective, we identify the following three preconditions for successful monitoring:

\begin{itemize}[leftmargin=*]

  \item \emph{Information availability: Do the oversight personnel have the right information?} The extent to which relevant cues are accessible, observable, and timely for human oversight personnel. Examples of relevant cues include deviations from expected output distributions, violations of domain-specific rules, valid confidence estimates, discrepancies between model outputs and valid external benchmarks, and indicators of mismatches between training and deployment contexts. Monitoring is degraded when relevant cues are hidden, delayed, or fragmented across systems. Availability is shaped by how information signals are captured and surface; e.g.,\ via system logs, telemetry data, uncertainty estimates, explanations, contextual input-output data access, and  alerts for anomalous behaviour. Temporal and spatial characteristics also shape availability: latency between cue occurrence and potential failure constrains timely intervention, and cues originating far from the point of failure may require clear mappings. Finally, signals may vary in frequency, modality, complexity, relevance, and completeness; they can be continuous streams or discrete. 

  \item \emph{Information detection: Can the oversight personnel detect the relevant cues?} The ability of human oversight personnel to notice and discriminate relevant cues. Detection can be impeded by poorly-designed interfaces, misaligned presentation of cues, or insufficient presentation time. Human attentional, motivational, and training deficits may further impair cue detection. Detection can be supported through clear interface designs, alerting and prioritisation mechanisms that manage information overload, and targeted training that focuses attention of relevance cues.

  \item \emph{Information interpretation: Do the oversight personnel interpret the cues correctly?} The process of integrating, weighting, and mapping cues to judge the likelihood and severity of problematic AI behaviour; ultimately informing intervention. Interpretation is shaped by prior beliefs, experience, human judgmental biases, technological support, and perceived costs of errors and interventions. Interpretation can be supported by defining clear standards for problematic behavior, training, documentation and records that support tracing, explanatory tools that clarify why cues are informative, and feedback on past outcomes.
\end{itemize}

\paragraph{Intervention}
Intervention is a process of deliberate planning and action. Particular types of interventions can be pre-determined---e.g.\ determined by policies, operational guidelines and standards ---or deliberated just-in-time--- e.g.\ informed by professional expertise of the oversight personnel. They can vary in scope and impact, and are shaped by the level of control, the nature of the task, and the characteristics of the intervention itself. Building on the work of \citet{sterz2024quest}, we define the following two preconditions for successful intervention:

\begin{itemize}[leftmargin=*]

  \item \emph{Oversight capability: Is oversight personnel able to act?} The oversight personnel's capacity to intervene, including the authority (i.e., whether they have been granted access or permissions), relevant skills/expertise (i.e., domain knowledge needed to execute the intervention), their willingness (i.e., tolerance for risks, perceived responsibility), their situation awareness (i.e., correct understanding of the situation, informed by monitoring), and their coordination in distributed oversight situations (e.g., when there are distributed responsibilities among the oversight personnel).
  
  \item \emph{Intervention feasibility: Is the intervention available and viable in the current context?} The extent to which potential intervention actions are available, applicable, and address the risk in the current situation. Feasibility varies by layer of oversight (see Figure \ref{fig:HO_two_layer_to_multi_layer_architecture}), and is shaped by characteristics of the interventions, such as frequency (i.e., how often intervention is required), effort (i.e., cognitive load, time, or physical resources needed to intervene), economic/reputational cost (i.e., economic implications of intervention, of delays), priority (i.e., time sensitivity or criticality), risk (i.e.,  potential harm or disruption of inaction or incorrect intervention), and governing policies and processes.
  
\end{itemize}

\section{A Template for Human Oversight in Practice}
\label{sec:template}


This section introduces a practical template for organising and evaluating human oversight of AI (see Appendix \ref{appendix:howto} for instructions on how to use the template and Appendices \ref{appendix:hiring}-\ref{appendix:diagnostics} for use cases). It was developed at the respective Dagstuhl Seminar \citep{Langer2026}, where participants used known requirements of effective oversight (e.g.\ \cite{sterz2024quest, langer2024effective}) to refine the framework from Section~\ref{sec:framework} into a set of key attributes required to implement oversight. It was then iteratively improved and verified in several case studies (see Section~\ref{sec:application}). 

The template operationalises our framework by providing a systematic way to document the oversight architectures and processes. It is intended for developers, deployers, auditors, regulators, and researchers, enabling them to think through oversight architectures and processes, document those, and also assess whether they are coherent, effective, and feasible.

The template could be used in participatory workshops, organisational audits, and design reviews, to surface design flaws, responsibilities, limitations, and dependencies. The resulting documentation supports risk management, compliance, and iterative improvement by translating abstract oversight principles into concrete, verifiable, and adaptable procedures. Note that the main aspects of the template are directly related to our oversight architecture and process, whereas the bullet points are examples for key aspects to consider when filling out the template.

\subsection{Task Layer}
The task layer represents the socio-technical system of the primary activity that involves the AI system. Human oversight focuses on this level’s functioning and its potential risks. 

\begin{longtable}{l@{~}p{0.12\textwidth}p{0.8\textwidth}}
\toprule
 & \textbf{Component} & \textbf{Description}\\
\midrule

1. & \textbf{Task} & Describe the object-level task to be performed: 
  \begin{itemize}[noitemsep,topsep=2pt,partopsep=2pt,leftmargin=*]
   \item \textit{primary goals}: intended outcomes of the task such as timely decision making.
   \item \textit{risks}: potential failures or harms associated with the task, such as safety hazards. 
   \item \textit{outcomes}: measurable results such as operational efficiency.
   \end{itemize} \\

2. & \textbf{AI system} & Describe the AI system that supports or automates parts of this task: 
   \begin{itemize}[noitemsep,topsep=2pt,partopsep=2pt,leftmargin=*]
   \item \textit{core properties}: main parts of the system architecture, such as models, platforms, etc.
   \item \textit{functionality}: what the system does, such as risk prediction.
   \item \textit{inputs and outputs}: data/information that it requires and produces.
   \item \textit{level of autonomy}: the degree of independence that the AI system has to perform functions without human intervention.
   \item \textit{known limitations} or \textit{failure modes}: conditions under which the system may underperform.
   \end{itemize}\\

3. & \textbf{Human(s)} & Identify the people directly interacting with, using, or affected by the AI system, such as operators or end-users. Describe their:
                       \begin{itemize}[noitemsep,topsep=2pt,partopsep=2pt,leftmargin=*]
                       \item \textit{abilities/competencies}: relevant skills and knowledge.
                       \item \textit{degree of discretion or authority}: their power to make decisions independently.
                       \item \textit{responsibilities}: duties and obligations assigned to the people.
                       \item \textit{intentions}: goals and priorities when using the system. 
                       \end{itemize} \\

4. & \textbf{Organisation} & Describe the organisational setting in which the task takes place, including:
                       \begin{itemize}[noitemsep,topsep=2pt,partopsep=2pt,leftmargin=*]
                       \item \textit{hierarchical structures}: reporting lines, responsibilities, team composition.
                       \item \textit{relevant rules, processes, and procedural constraints}: standard operating procedures, compliance requirements, resource availabilities.
                       \end{itemize}\\
\bottomrule
\end{longtable}

\subsection{Oversight Layer}
The purpose of the oversight layer is to monitor the operation of the AI system in context, to detect risks or failures, and to intervene when necessary. Oversight at this level ensures that the AI system functions as intended, including safety, alignment with regulatory requirements, norms, and standards. 

For each relevant layer (i.e. real-time, systemic, and compliance layers), define:

\begin{longtable}{l@{~}p{0.12\textwidth}p{0.8\textwidth}}
\toprule
 & \textbf{Component} & \textbf{Description}\\
\midrule
1. & \textbf{Oversight Task} & Describe the oversight task to be performed: 
  \begin{itemize}[noitemsep,topsep=2pt,partopsep=2pt,leftmargin=*]
   \item \textit{purpose}: why oversight is needed, e.g. ensuring safety.
   \item \textit{goals}: specific objectives that oversight aims to achieve, such as detecting anomalies.
   \item \textit{outcomes}: measurable results of the oversight activity, such as reduced error rates.
   \item \textit{risks}: potential harms associated with the oversight task, such as unclear escalation.
   \end{itemize}\\

2. & \textbf{Oversight Technology (Optional)} & Describe the technologies that support oversight, such as monitoring dashboards, warning systems, or audit logs. Describe their:
   \begin{itemize}[noitemsep,topsep=2pt,partopsep=2pt,leftmargin=*]
   \item \textit{properties}: key characteristics such as real-time monitoring capability.
   \item \textit{functionalities}: what the system does, such as display performance metrics.
   \item \textit{failure modes}: conditions under which these tools might fail, such as data latency.
   \item \textit{system interfaces}: how these tools interface with the AI system, such as API interaction.
   \item \textit{signals}: how they communicate relevant information to the human overseers, such as visual alerts, reports etc.
   \end{itemize}\\

3. & \textbf{Oversight Personnel} & Describe the personnel responsible for oversight, including:
   \begin{itemize}[noitemsep,topsep=2pt,partopsep=2pt,leftmargin=*]
   \item \textit{abilities/competencies}: knowledge, skills, abilities required, such as technical expertise.
   \item \textit{authority}: decision-making power, such as the ability to halt the system.
   \item \textit{responsibility}: tasks overseers must perform, such as monitoring and intervention duties, decision thresholds, and escalation procedures.
   \item \textit{intentions}: goals and priorities guiding actions, such as complying with regulations.
   \end{itemize} \\

4. & \textbf{Oversight Organisation} & Describe the organisational setting in which the oversight takes place, including:
   \begin{itemize}[noitemsep,topsep=2pt,partopsep=2pt,leftmargin=*]
    \item \textit{composition of oversight}: individual or team oversight; single organisation or distributed oversight.
    \item \textit{rules and procedural constraints}: standard operating procedures and audit protocols;
   \item \textit{governance structures}: committees or bodies allocating oversight responsibilities.
   \item \textit{reporting mechanisms}: incident logs, escalation pathways.
   \item \textit{coordination}: how oversight activities align with other teams (IT, legal, operations).
   \item \textit{constraints}: resource limitations or technological dependencies affecting feasibility.
   \end{itemize} \\
\bottomrule
\end{longtable}

\subsection{Oversight Process}

The oversight process describes the relevant activities in the oversight task. For each of the real-time, systemic, and compliance layers, define:

\begin{longtable}{l@{~}p{0.12\textwidth}p{0.8\textwidth}}
\toprule
 & \textbf{Component} & \textbf{Description}\\
\midrule

1. & \textbf{Monitor} & Specify the types of monitoring activities and their temporal frequency:
   \begin{itemize}[noitemsep,topsep=2pt,partopsep=2pt,leftmargin=*]
   \item \textit{signals}: the signals received from the overseen layer.
   \item \textit{interpretation}: the process for interpreting signals.
   \item \textit{frequency}: how often monitoring occurs, such as continuous/real-time, scheduled, or conditional on other events.
   \end{itemize}\\

2. & \textbf{Assess Risk} & Define the risk assessment process to be used by the oversight personnel.\\

3. & \textbf{Decide} & Define the decision-making process that the oversight personnel follow, including: 
   \begin{itemize}[noitemsep,topsep=2pt,partopsep=2pt,leftmargin=*]
   \item \textit{decision protocols}: predefined rules or human judgment.
   \item \textit{documentation}: record rationale for accountability and audit purposes.
   \end{itemize}\\

4. & \textbf{Intervene} & Specify intervention types and their temporal frequency (real-time, scheduled, or conditional).\\
\bottomrule
\end{longtable}

\subsection{Application, Evaluation, and Insights}
\label{sec:application}

We applied the template to four domains: hiring, air traffic control, automated insulin delivery, and medical diagnostics, to test feasibility and practical value (Appendices~\ref{appendix:hiring}-\ref{appendix:diagnostics}). In all cases, the template supported a structured reflection on the scope and purpose of oversight, clarified who bears responsibilities, and identified the institutional and technical supports required. For example, in the hiring case, documenting roles and decision thresholds revealed a risk of diffusion of responsibility across two formally designated overseers, motivating a need for clearer role definitions and accountability mechanisms. Completing the template surfaced dependencies and risks that otherwise might have been overlooked. 

While the template is generalisable, implementation necessarily reflects contextual factors such as regulatory obligations, technical maturity, and organisational culture. We also saw that concrete and in-depth responses to the template require integrating perspectives from individuals with varied domain expertise (e.g., individuals who work directly with the AI-based system on a task), technical expertise, and AI governance and ethics expertise.

\section{Challenges Facing Effective Human Oversight}
\label{sec:futurework}


Although our framework provides a practical starting point for implementing effective human oversight of AI, it also exposes central challenges: \textit{operational challenges} in the core monitoring and intervention activities; \textit{socio-technical challenges} that consider the interactions between the human, the task and the organisational context; and \textit{governance challenges}, concerning the limits and legitimacy of human oversight.

\subsection{Operational Challenges}
Effective oversight depends on humans being able to \textit{monitor} AI systems and appropriately \textit{intervene} when needed, activities that pose difficulties that go beyond those encountered in traditional human-automation interaction. Moreover, human overseers may require new forms of expertise to meaningfully enact their responsibilities.

 \textit{Monitoring} AI systems often involves interpreting probabilistic, multi-modal, and dynamically changing information signals rather than deterministic ones~\citep{naveed2025monitoring}. Signals may drift over time, emerge only at aggregate levels, or be distributed across system components \citep{naveed2025monitoring,shankar2024we, protschky2025monitoring}. A challenge, therefore, is determining which information signals should be provided and how they should be represented across different task layers, risk profiles, and oversight roles.  

\textit{Intervening} in AI systems can require actions at multiple layers, including model modifications, full retraining, changes to data pipelines, or the introduction of human-in-the-loop validation \citep{shankar2024we,bhargava2024challenges}. Effective intervention, therefore, requires mapping how AI systems are developed, deployed, and maintained across data sources, supply chains, and model layers ~\citep{sambasivan2021everyone,balayn2025unpacking}.

\subsection{Socio-technical Challenges}
Research in human-automation interaction has identified persistent factors that impede human control \citep{Kaber2025,mosier2019humans,sheridan2021chapter}. Fundamentally, human oversight is constrained by well-documented limits of human cognition, including reliance on heuristics (i.e., bounded rationality, \citep{simon1990bounded,mosier2019humans,hollan2000distributed}). These limits are compounded by the restricted effectiveness of explainability \citep{kaur2022sensible,miller2023explainable,he2023knowing,rieger2023careful,gaube2023non,gaube2026underreliance,cecil2024explainability,tintarev2025} and transparency \citep{liao2023designerly, hussain2025transparency}, which do not guarantee meaningful understanding or reliable intervention.
Oversight tasks also frequently involve prolonged periods of passive monitoring, a mode of engagement for which humans are poorly suited due to limitations in sustaining vigilance \citep{klein2025vigilance}.
These challenges may be further exacerbated by organisational goals and incentive structures that undermine effective oversight, such as productivity targets that reward rapid acceptance of AI outputs rather than critical engagement \citep{shankar2024we,parasuraman2010complacency}. 

Even where human oversight is feasible in principle, sustaining it in practice requires carefully designed socio-technical conditions \citep{cummings2023frontiers,grote2023,naikar2023,pritchett2024things}. Effective oversight depends on aligning interfaces, workflows, organisational structures, and training with the demands of monitoring and intervention. 
Oversight also increasingly requires AI-specific technical competencies, including understanding model behaviour and error modes \citep{naveed2025monitoring}. These demands intensify in multi-model and multi-agent systems, where interacting components can give rise to emergent and difficult-to-interpret behaviours, and thus understanding how to exert human control and oversight becomes an even more complex challenge \citep{allmendinger2026multi}. For such systems, the kind of real-time oversight that this paper mainly focuses on may be infeasible \citep{passi2025agentic}. They might additionally require containment architectures that enable anticipatory oversight by providing systems with guardrails before deployment \citep{jahn2026breaking, wood_2025autonomous}. 

At the same time, increasing reliance on AI introduces a central dilemma: using AI systems may undermine the expertise needed to oversee them. Cognitive offloading, automation bias, reduced practice, and missing feedback loops contribute to deskilling and erosion of expertise, even in safety-critical settings \citep{rinta2023vicious,budzyn2025endoscopist}. As AI systems become more capable, humans may have fewer opportunities or incentives to develop and maintain domain expertise \citep{endsley2023ironies, Bainbridge_1983}. Sustaining human oversight, therefore, requires deliberate investment in training and long-term capability maintenance of the oversight personnel -- and foundational discussions on where we want to invest time and resources to maintain these capabilities.

\subsection{Governance Challenges }
When the challenges outlined above are not adequately addressed, human oversight risks becoming symbolic rather than effective or meaningful. The mere presence of oversight personnel or structures does not guarantee risk mitigation and may instead create an illusion of control \citep{green2022flaws}. In some cases, responsibility for failures is displaced from designers or organisations onto individual overseers, who become scapegoats \citep{elish2019moral}. Finally, scholars have questioned whether meaningful human oversight is ethically defensible in some domains (e.g., in the use of lethal autonomous weapon systems \citep{schwarz2021autonomous}), and in other contexts, challenges for privacy may arise (e.g., when human oversight is supposed to mitigate risks in user-AI interactions in healthcare). Addressing this challenge requires establishing methods for evaluating and monitoring oversight effectiveness and clear definitions of roles, responsibilities, and evaluation criteria. However, testing whether human oversight actually meets regulatory requirements poses its own challenges, as compliance criteria are difficult to operationalise and standard evaluation methods may be insufficient for the socio-technical complexity of oversight \citep{Langer2026,laux2025automation}. Moreover, oversight architectures themselves can become targets for adversarial exploitation, introducing a further layer of risk \citep{ditz2025secure}.

\subsection{Overall} These challenges prompt key questions:
 \begin{enumerate}
     \item  \textit{Operational}: What monitoring and intervention capabilities must be available for oversight to be meaningful and effective?
     \item  \textit{Socio-technical}: Under what conditions is it reasonable to assign oversight responsibility to humans, and how should feasible oversight goals be set? How can oversight-related expertise, as well as technical and organisational support be designed and sustained over time?
     \item  \textit{Governance}: How to evaluate and monitor the effectiveness of human oversight and what normative, regulatory, or design responses are warranted in cases where meaningful oversight is not achievable?
 \end{enumerate}

\section{Conclusion}



In this paper, we proposed a framework for effective human oversight of AI systems. Built on cross-disciplinary perspectives from participants at the 2025 Dagstuhl Seminar on human oversight, the key insights reflected in the framework are: \textbf{(a) }that human oversight of AI is a multi-layered issue, which goes beyond real-time intervention to include systemic and compliance layers; and \textbf{(b)} monitoring and interventions are required at these layers, and require new areas of expertise than we have seen with oversight of `traditional automation'. We provided a template for human oversight architectures and processes designed to engage stakeholders when reflecting on the socio-technical conditions for effective human oversight. We have shown this template can be applied in diverse domains, demonstrating the practical applicability of the framework. 

We hope that this framework provides a shared foundation for researchers and practitioners working on human oversight of AI systems. However, it cannot resolve all challenges that arise as AI systems, especially systems with agentic capabilities that raise new practical, ethical, and legal demands for human control. We have, therefore, outlined key challenges for future research on human oversight, to focus the community's research efforts. Many of the issues raised are not completely new, but they are arguably more complex than those seen before. Furthermore, the growing capabilities and widespread deployment of AI systems render these issues as urgent, and a major research program on effective human oversight of AI will be required to resolve them.

\subsection*{Acknowledgements}

The conceptual analysis for this work was undertaken by the authors at Dagstuhl seminar 25272 \emph{Challenges of Human Oversight: Achieving Human Control of AI-Based Systems} \url{https://www.dagstuhl.de/25272} \cite{langer_challenges_2026}, held at Schloss Dagstuhl (June 29th-July 4th, 2025).

Individual authors acknowledge the support of:
Kevin Baum, Raimund Dachselt, Markus Langer, and Hanwei Zhang by the German Research Foundation (DFG) under grant No. 389792660, as part of TRR 248, see https://perspicuous-computing.science, Kevin Baum by the German Federal Ministry of Education and Research (BMBF) as part of the project MAC-MERLin (Grant Agreement No. 16IW24007), and by the European Regional Development Fund (ERDF) and the Saarland within the scope of (To)CERTAIN (Project ID EFRE-AuF-0000942).
Markus Langer by the Daimler and Benz Foundation as part of the project TITAN – Technologische Intelligenz zur Transformation, Automatisierung und Nutzerorientierung des Justizsystems (grant no. 45-06/24). Ujwal Gadiraju was partially supported by the TU Delft AI Initiative, the Model-driven Decisions Lab, the GENIUS ICAI Lab, and the Convergence Flagship project \textit{ProtectMe.} Mark Keane by Research Ireland (Taighde Éireann) through the Insight Centre for Data Analytics (12/RC/2289 P2). Hanwei Zhang by DFG project \textit{CARAT} (grant 547583482). Susanne Gaube was supported by \textit{UCL Grand Challenges}, and\textit{ UCL Global Engagement} initiatives, and by the European Commission Marie Skłodowska-Curie Actions Programme as part of the project \textit{GRASP} (\#GAP-101236769). Nava Tintarev  was supported by the project ROBUST: Trustworthy AI-based Systems for Sustainable Growth with project number KICH3.LTP.20.006, which is (partly) financed by the Dutch Research Council (NWO), RTL, DPG, and the Dutch Ministry of Economic Affairs and Climate Policy (EZK) under the program LTP KIC 2020-2024. Johann Laux was supported by a British Academy Postdoctoral Fellowship (grant no. PF22\textbackslash{}220076) for the project "The Emerging Laws of Oversight". Brian Lim was supported by he National Research Foundation, Singapore and Infocomm Media Development Authority under its Trust Tech Funding Initiative (Award No. DTC-RGC-09).
All content represents the opinion of the authors, which is not necessarily shared or endorsed by their
respective employers and/or sponsors. The authors would like to thank Laura Stenzel and  Susan Shelmerdine for additional comments.

\bibliographystyle{plainnat}
\bibliography{references}

\appendix 
\section{How to Use the Documentation Template}\label{appendix:howto}

\subsection{Note on the Appendices B-E}
Appendices \ref{appendix:hiring}–\ref{appendix:diagnostics} present documentation of human oversight across a range of use cases, prepared by sub-teams of co-authors with domain-specific expertise. We deliberately preserved variation in how the templates were completed---that is, we did not post-edit for consistency in the sub-points addressed, the layers of human oversight considered, or the level of detail provided. This exercise is to illustrate that the template can be flexibly adapted to the specific requirements of different application domains.

\subsection{How to Use the Documentation Template}
The following steps provide a practical guide for applying the human oversight framework to a specific AI use case. This template is designed to capture a broader range of oversight processes than traditional models; however, it is intended to be flexible. Users are not required to address every layer or bullet point if they are not relevant to their specific context or organisational responsibility.

\begin{enumerate}
    \item \textbf{Define the Scope of Integration:} 
    Identify which oversight layers are relevant to your project or role. For example, if you are a system designer, you might focus on the \textit{Object-Level Task} and \textit{Real-Time Oversight} layers. If you are developing organisational policy, you might prioritise the \textit{Object-Level Task} and \textit{Compliance} layers.
    
    \item \textbf{Identify Areas of Responsibility:} 
    Determine for which specific layers you or your organisation hold primary responsibility. While the framework presents a holistic view, your documentation should prioritise the processes within your direct sphere of influence or authority.
    
    \item \textbf{Populate the Layer Templates:} 
    Fill out the descriptive sections for your chosen layers. You should treat the provided bullet points (e.g., goals, risks, technology properties) as illustrative examples rather than an exhaustive checklist. Feel free to modify, add, or omit points to best reflect your domain-specific needs.
    
    \item \textbf{Detail the Oversight Processes:} 
    Using the information established in the previous step, complete the \textit{Oversight Process} template for each included layer. This involves specifying the "Monitor—Assess Risk—Decide—Intervene" loop specific to that layer's temporal frequency and goals.
\end{enumerate}

\noindent \textit{Note: The goal of this documentation is to surface design flaws and clarify dependencies. Focus on the layers that provide the most value for risk mitigation in your specific context.}

\section{Hiring}\label{appendix:hiring}

\subsection{Task Layer}
This appendix outlines an example of the human oversight framework applied to the domain of hiring where an AI system assesses the fit of applicants against job requirements based on recorded job interviews.

\begin{longtable}{l@{~}p{0.12\textwidth}p{0.8\textwidth}}

1. & \textbf{Object-Level Task} & Describe the object-level task to be performed: 
  \begin{itemize}[noitemsep,topsep=2pt,partopsep=2pt,leftmargin=*]
   \item \textit{primary goals}: The task is to assess applicant fit against job requirements.
   \item \textit{risks}: unfair discrimination of individual applicants or groups of applicants, i.e., applicant fit is judged to be lower than the respective applicants' actual suitability.
   \item \textit{outcomes}: Time-to-hire; job performance of hired individuals in the future.
   \end{itemize} \\

2. & \textbf{AI system} & Describe the AI system that supports or automates parts of this task: 
   \begin{itemize}[noitemsep,topsep=2pt,partopsep=2pt,leftmargin=*]
   \item \textit{core properties}: The AI system is an automated job interview evaluation system. Applicants record videos of themselves responding to interview questions. There is a transcription module that transcribes applicant responses and an evaluation module that was trained on 1000 prior applicants and their responses.
   \item \textit{functionality}: The AI automatically assesses applicant suitability from 0 to 100 percent fit.
   \item \textit{inputs and outputs}: Input: applicant videos; Output: applicant fit score.
   \item \textit{level of autonomy}: The AI fully automates the applicant evaluation.
   \item \textit{known limitations} or \textit{failure modes}: Low audio quality can undermine transcription accuracy; low audio quality paired with applicants whose first language is not English leads to even worse transcription accuracies which can reduce applicant suitability ratings.
   \end{itemize} \\

3. & \textbf{Human(s)} & Identify the people directly interacting with, using, or affected by the AI system, such as operators or end-users. Describe their:
   \begin{itemize}[noitemsep,topsep=2pt,partopsep=2pt,leftmargin=*]
   \item \textit{abilities/competencies}: The operator is a hiring manager trained in assessing applicant fit. They have also received some training in the use of the AI system, e.g., on how to interpret the applicant suitability score.
   \item \textit{degree of discretion or authority}: The operator uses the outputs of the AI system for deciding who to invite to an in-person interview. They can also ignore the AI output for this task.
   \item \textit{responsibilities}: They are responsible for evaluating applicants, for ensuring that hiring processes follow legal standards, e.g., of non-discrimination. They also communicate with invited and rejected applicants.
   \end{itemize} \\

4. & \textbf{Organisation} & Describe the organisational setting in which the task takes place, including:
   \begin{itemize}[noitemsep,topsep=2pt,partopsep=2pt,leftmargin=*]
   \item \textit{hierarchical structures}: The hiring manager is responsible for the successful and quick task completion, and they report to their group leader.
   \item \textit{relevant rules and procedural constraints}: The organization has a human resources group that defines the hiring processes.
   \end{itemize}\\

\end{longtable}

\subsection{Oversight Layer}
The purpose of the oversight layer is to monitor the operation of the AI system in context, to detect risks or failures, and to intervene when necessary. Oversight at this level ensures that the AI system functions safely, fairly, and as intended. 

\begin{longtable}{l@{~}p{0.22\textwidth}p{0.73\textwidth}}

1. & \textbf{Oversight Task} & Describe the oversight task to be performed: 
  \begin{itemize}[noitemsep,topsep=2pt,partopsep=2pt,leftmargin=*]
   \item \textit{purpose}: The evaluation AI may provide inaccurate results and under- or overestimate the actual applicant suitability especially when audio quality of applicant interviews is low. The evaluation AI may also produce discriminatory outputs if applicants do not have English as their first language.
   \item \textit{goals}: 1. Assess the adequacy of the AI evaluation module for individual applicants; 2. Assess whether the AI evaluation module is functioning properly on a group level; and 3. Assess whether the AI evaluation module provides discriminatory outputs.
   \item \textit{risks}: Inadequacy of using the AI evaluation may remain undetected, for instance, due to the large number of applicants applying and due to the limited time provided for real-time oversight.
   \end{itemize}\\

2. & \textbf{Oversight Technology (Optional)} & Describe the technologies that support oversight, such as monitoring dashboards, warning systems, or audit logs. Describe their:
   \begin{itemize}[noitemsep,topsep=2pt,partopsep=2pt,leftmargin=*]
   \item \textit{functionalities}: Real-time oversight (i.e., hiring managers) has access to a warning tool that monitors word error rates in transcriptions and audio quality. 
   
   Systemic oversight (i.e., technicians) has access to the benchmark dashboard that includes warning modules that highlight benchmarks that are out of bounds and that assess different fairness metrics.
   \end{itemize}\\

3. & \textbf{Oversight Personnel} & Describe the personnel responsible for oversight, including:
   \begin{itemize}[noitemsep,topsep=2pt,partopsep=2pt,leftmargin=*]
   \item \textit{composition}: Real-time oversight personnel: a hiring manager overseeing AI use for individual applicants.
   Systemic oversight personnel: a technician overseeing the aggregate functioning of the AI system.
   Compliance oversight personnel: compliance officers assessing whether AI behavior reflects fairness issues.
   \item \textit{abilities/competencies}: Real-time oversight personnel is trained in assessing applicant fit and in using the AI system as decision support. 
   Systemic oversight personnel is trained in assessing technical AI benchmarks.
   Compliance oversight personnel is trained in assessing AI fairness challenges.
   \item \textit{responsibility}: Real-time oversight: Their duty is to assess for each individual applicant whether the AI evaluation system is adequate to use. 
   Systemic oversight: Their duty is to assess on a group level whether the system is functioning adequately. They monitor whether AI benchmarks are critically out of bounds.
   Compliance oversight: Their duty is to assess whether reported challenges reflect fairness issues.
   \end{itemize} \\

4. & \textbf{Oversight Organisation} & Describe the organisational setting in which the oversight takes place, including:
   \begin{itemize}[noitemsep,topsep=2pt,partopsep=2pt,leftmargin=*]
   \item \textit{reporting mechanisms}: Real-time oversight reports to the systemic- and compliance oversights if they detect issues for individual applicants.
   Systemic oversight reports to the real-time and to compliance oversight personnel if they detect issues on a group level. Compliance oversight reports back to the other layers if critical fairness challenges arise.
   \end{itemize} \\


\end{longtable}

\subsection{Oversight Process}

The oversight process describes all the relevant activities in the oversight task. 

\begin{longtable}{l@{~}p{0.16\textwidth}p{0.76\textwidth}}

1. & \textbf{Monitor} & Specify the types of monitoring activities and their temporal frequency:
   \begin{itemize}[noitemsep,topsep=2pt,partopsep=2pt,leftmargin=*]
   \item \textit{signals}: Real-time oversight monitoring duties: Examine the word error rate for single individuals, check for audio quality issues in videos, check for accents and dialects that may undermine the quality of the AI evaluation module. 
   Systemic oversight monitoring duties: monitor benchmarks over time such as shifts in suitability ratings depending on different subgroups of applicants.
   Compliance oversight monitoring duties: monitor and assess whether reported issues indeed represent fairness issues.
   \end{itemize} \\

2. & \textbf{Assess Risk} & Define the risk assessment process to be used by the oversight personnel:
    \begin{itemize}[noitemsep,topsep=2pt,partopsep=2pt,leftmargin=*]
    \item The main risk is that the AI evaluation module is inadequate to assess specific applicants due to the transcription module producing higher error rates for certain applicants and certain technical circumstances.
    \end{itemize} \\

3. & \textbf{Decide} & Define the decision making process that the oversight personnel follow, including: 
   \begin{itemize}[noitemsep,topsep=2pt,partopsep=2pt,leftmargin=*]
   \item \textit{decision protocols}: All oversight personnel is instructed to decide for interventions based on their respective oversight training, duties and based on their own discretion.
   \item \textit{documentation}: Oversight personnel is instructed to document by reporting to the other oversight layers.
   \end{itemize} \\

4. & \textbf{Intervene} & Specify intervention types and their temporal frequency (real-time, scheduled, or conditional).
   \begin{itemize}[noitemsep,topsep=2pt,partopsep=2pt,leftmargin=*]
   \item Real-time oversight: Ignore AI evaluation module and assess applicant suitability independently.
   \item Systemic oversight: Provide a warning signal to real-time and compliance oversight; stop the AI evaluation module.
   \item Compliance oversight: Provide a warning signal to the systemic and real-time oversight, stop the AI evaluation module and suggest solutions to fix fairness issues.
   \end{itemize} \\

\end{longtable}

\section{Air Traffic Control}

This appendix outlines an example of the human oversight framework applied to the domain of air traffic control (ATC) --- a part of air traffic management that involves real-time guidance to aircraft from take-off, flying, and landing.

\subsection{Task Layer}

The task layer consists of \emph{teams} of air-traffic controllers, each of which controls a particular sector of airspace, with managers/supervisors who help to monitor air traffic at a higher level, considering factors such as traffic flow and monitoring for emergencies. Air traffic controllers and managers are assisted by various technological tools, from tasks such as automated aircraft tracking, to collision detection, to decision support, such as which aircraft should change its flight path and how in case of a potential collision.

\begin{longtable}{l@{~}p{0.17\textwidth}p{0.78\textwidth}}

1. & \textbf{Object-Level Task} & Describe the object-level task to be performed:
  \begin{itemize}[noitemsep,topsep=2pt,partopsep=2pt,leftmargin=*]
   \item \textit{primary goals}: The goals of object-level task include allocating aircraft to airspace to ensure the safety of people on board and on the ground, while also ensuring a smooth flow of air traffic (e.g. avoiding delays). Individually, controllers allocate paths to aircraft within their own air sector, but are also responsible for the safe handover of aircraft between their sector and that of other controllers. The task is accomplished through a distributed system which spans multiple controllers, interactive artifacts (e.g., flight strips, radar), and organisational structure.
   \item \textit{risks}: The main risks are: potential accidents, and  delays in flights, with the constraint that this is a real-time scheduling problem that must be solved "in flight"---aircraft cannot be parked or paused.
   \item \textit{outcomes}: The outcomes are whether aircraft land safely and on time, and a successful transformation of the continuous flow of air traffic into a manageable sequence.
   \end{itemize} \\

2. & \textbf{AI system} & Describe the AI system that supports or automates parts of this task:
   \begin{itemize}[noitemsep,topsep=2pt,partopsep=2pt,leftmargin=*]
   \item \textit{core properties}: There are several technological components in ATC, but some critical examples using AI include prediction of air traffic 20-30 minutes in advance, prediction of aircraft trajectory, collision prediction, and collision resolution (recommending conflict-free trajectories).
   \item \textit{inputs and outputs}: inputs such as flight schedules, information on delays, information on aircraft; outputs: prediction of aircraft trajectory. 
   \item \textit{level of autonomy}: The AI can work completely autonomously, such that controllers do not have to rely on mental simulations of flight trajectories.
   \item \textit{known limitations} or \textit{failure modes}: Failure can result when incorrect information is provided to the AI, when predictions of traffic, trajectories, and collisions are incorrect, or when collision resolution fails; e.g., solves one conflict but introduces another, or does not have access to necessary information that the controller has, including controller preferences. Moreover, automating some tasks can disrupt the working division of labor if it does not account for how different stakeholders coordinate through shared artifacts that they use to make decisions.
   \end{itemize} \\

3. & \textbf{Human(s)} & Identify the people directly interacting with, using, or affected by the AI system, such as operators or end-users. Describe their: 
   \begin{itemize}[noitemsep,topsep=2pt,partopsep=2pt,leftmargin=*]
   \item \textit{abilities/competencies}: Air traffic controllers, whose decisions are in real time and can have major safety and air traffic flow consequences. These are highly-trained professionals in the core task of air traffic management. They have also received some training on the AI system, for instance regarding its known limitations.
   \item \textit{degree of discretion or authority}: They have the ability to change flight departure time, arrival time, and flight paths, and the authority to tell pilots to do these.
   \item \textit{responsibilities}: They are responsible for ensuring safety and air traffic flow.
   \end{itemize} \\

4. & \textbf{Organisation} & Describe the organisational setting in which the task takes place, including: 
   \begin{itemize}[noitemsep,topsep=2pt,partopsep=2pt,leftmargin=*]
   \item \textit{hierarchical structures}: The organisations involved include airlines, air services, and airports. Air services have authority to dictate---to airlines and airports---flight departure, arrival, paths, etc. Pilots and air traffic controllers report regularly to each other on flight updates.
   \item \textit{relevant rules and procedural constraints}: Procedurally, aviation is one of the most heavily regulated and compliancy-based sectors.
   \end{itemize}\\

\end{longtable}

\subsection{Oversight Layer}

\subsubsection{Real-time layer}

The real-time oversight layer helps to ensure safety and air traffic flow by monitoring and intervening on both ATC systems and controllers, in real-time:

\begin{longtable}{l@{~}p{0.22\textwidth}p{0.73\textwidth}}

1. & \textbf{Oversight Task} & Describe the oversight task to be performed:
  \begin{itemize}[noitemsep,topsep=2pt,partopsep=2pt,leftmargin=*]
   \item \textit{purpose and goals}: The primary oversight task is to ensure safety by monitoring AI system outputs (trajectory predictions, conflict alerts, resolution recommendations) and intervening when these systems provide inaccurate, incomplete, or unsafe guidance. Secondary goals include maintaining efficient traffic flow and ensuring regulatory compliance. Oversight is not a discrete check but is woven into the ongoing work of controlling: monitoring occurs continuously as controllers interact with AI-generated predictions and recommendations; each decision to accept or override an AI suggestion constitutes human oversight activity.
   \item \textit{risks}: The oversight may fail due to unclear responsibilities between the ATC and their supervisor; the oversight may fail due to the significant workload for ATCs.
   \end{itemize}\\

2. & \textbf{Oversight Technology (Optional)} & Describe the technologies that support oversight, such as monitoring dashboards, warning systems, or audit logs. Describe their: 
   \begin{itemize}[noitemsep,topsep=2pt,partopsep=2pt,leftmargin=*]
   \item \textit{functionalities}: The real-time oversight technology include warning systems for potential collisions or congestion, load-balancing dashboards for sector merging for supervisors to monitor sector loads per pilot, intervene in high- or low-load sectors by merging contiguous sectors, or splitting sectors. These oversee the great socio-technical system, not just the software. These tools function as cognitive artifacts that off-load some monitoring burden (e.g., automated scanning for conflicts) while introducing new coordination demands (e.g., understanding why a conflict alert was triggered, distinguishing true alerts from false positives).
   \item \textit{system interfaces}: This oversight technology interfaces with AI systems by receiving real-time predictions and comparing them against actual aircraft behavior captured by radar and data links.
   \item \textit{signals}: Warning systems for human overseer(s) may use visual (color-coded alerts, flashing indicators) and auditory signals to communicate urgency.
   \end{itemize}\\

3. & \textbf{Oversight Personnel} & Describe the personnel responsible for oversight, including:
   \begin{itemize}[noitemsep,topsep=2pt,partopsep=2pt,leftmargin=*]
   \item \textit{composition}: The two main oversight personnel are the front-line controllers and their supervisors. Oversight responsibility is distributed across multiple roles operating at different timescales and hierarchies.
   \item \textit{abilities/competencies}: Front-line controllers handle real-time oversight through continuous interaction with the AI task systems. They possess deep expertise in traffic patterns, aircraft performance characteristics, local airspace constraints, and weather impacts. Supervisors maintain shift-level oversight by monitoring system performance at an aggregate level across multiple sectors, tracking both automated tools and controller workload. They have expertise in system behavior patterns, organisational coordination, and resource allocation.
   \item \textit{authority}: Controllers and supervisors have authority to immediately override or ignore AI recommendations in their sector. They hold authority to disable any malfunctioning systems, reallocate workload, and escalate issues to on-call or offline technical teams.
   \item \textit{responsibility}: Front-line controllers are responsible for monitoring the air-traffic control systems and intervening in situations that have the potential to cause an incident within their sector or at the boundary between their airspace and a neighbouring sector. Supervisors are expected to identify potential conflict \emph{across} different airspaces and ensuring even spread of traffic at an aggregate level across multiple sectors, tracking both automated tools and controller workload.
   \end{itemize} \\

4. & \textbf{Oversight Organisation} & As above.\\

\end{longtable}

\subsubsection{Systemic layer}

\begin{longtable}{l@{~}p{0.22\textwidth}p{0.73\textwidth}}

1. & \textbf{Oversight Task} & Describe the object-level task to be performed:
  \begin{itemize}[noitemsep,topsep=2pt,partopsep=2pt,leftmargin=*]
   \item \textit{purpose and goals}: The systemic oversight task considers factors beyond the final predictions (e.g., of trajectories, load balancing) to intermediate computational steps that would enable auditing. For example, for trajectory prediction, this may be done by logging the aircraft performance, wind estimates used, waypoint calculations, etc. -- such that domain experts can use a model's explanatory output to trace why a prediction was (in)accurate. This level of oversight must also be responsive to real-time alternatives (e.g., through counterfactual predictions); multiple reports (e.g., when controllers in different sectors encounter similar problems); and added situational factors (e.g., updating the model weights to prioritize weather information). These intermediate computational steps also contextualise the design task for AI system developers who  work with the controllers to make updates to the system.
  \end{itemize} \\

2. & \textbf{Oversight Technology} & Describe the technologies that support oversight, such as monitoring dashboards, warning systems, or audit logs. Describe their:

   \begin{itemize}[noitemsep,topsep=2pt,partopsep=2pt,leftmargin=*]
    \item \textit{functionalities}: These would typically be immutable logging systems for traceability of runtime decisions, tools for aggregating individual flight information, and replay tools to replay incidents that occur, as well as tools for safety audit, such as root-cause analysis and interpretability tools.
   \end{itemize} \\

3. & \textbf{Oversight Personnel} & Describe the personnel responsible for oversight, including: 
   \begin{itemize}[noitemsep,topsep=2pt,partopsep=2pt,leftmargin=*]
    \item \textit{composition}: Experienced controllers (e.g., supervisors) continually test AI performance, coordinate across sector controllers to identify common themes of AI errors, and make decisions about systematic AI use in different scenarios. System and safety engineers (e.g., at Air Navigation Service Providers) bridge operations and development. AI developers and data scientists analyze logged data (e.g., accepted/rejected recommendations, computational trace data) to identify potential algorithmic failures and retrain models.
    \item \textit{abilities/competencies}: For effective oversight, these may be controllers with technical training, which gives them a combination of prior domain expertise and acquired technical skills to make changes. This requires expertise in AI and software development, supported by design technology such as explanatory outputs.
    \item \textit{responsibility}: Experienced controllers (e.g., supervisors) continually test AI performance, coordinate across sector controllers to identify common themes of AI errors, and make decisions about systematic AI use in different scenarios. System and safety engineers (e.g., at Air Navigation Service Providers) bridge operations and development: they diagnose root causes of controller-reported issues, translating operational problems into technical specifications. AI developers and data scientists analyze logged data (e.g., accepted/rejected recommendations, computational trace data) to identify potential algorithmic failures and retrain models.
   \end{itemize} \\

4. & \textbf{Organisation} & Describe the organisational setting in which the oversight takes place, including: assessment and support technology.
   \begin{itemize}[noitemsep,topsep=2pt,partopsep=2pt,leftmargin=*]
    \item \textit{rules and procedural constraints}: Regulatory bodies set formal reporting and performance guidelines (as above). Aviation software certification standards are rigid due to the safety-critical system needs. They require formal verification, hazard analysis, and extensive safety documentation that can extend design timelines. Legacy infrastructure dependencies can be an issue, which require modern AI systems to interface with decades-old radar, communication, and data processing and presentation systems.
    \item \textit{governance structures}: These are operationalized as hierarchies within the ATC teams and supervisors, with different tasks for local vs. global AI assessment and support technology.
   \end{itemize} \\



\end{longtable}

\subsubsection{Compliance Layer}


\begin{longtable}{l@{~}p{0.12\textwidth}p{0.8\textwidth}}

1. & \textbf{Oversight Task} & Describe the oversight task to be performed: 
  \begin{itemize}[noitemsep,topsep=2pt,partopsep=2pt,leftmargin=*]
   \item \textit{purpose}: Assurance that an organisation is complying with external and internal regulations in air-traffic control.
   \item \textit{primary goals}: Overseers are typically tasked with the goal of demonstrating and maintaining compliance  to safety regulation, as well as ensuring air traffic flow requirements set internally by their organisation.
    \item \textit{risks}: The risks are that they fail to monitor compliance appropriately, misinterpret regulation, and do not have sufficient information to identify when to intervene.
   \item \textit{outcomes}: Verified compliance, correct interventions such as requesting software improvements.

  \end{itemize} \\

2. & \textbf{Oversight Technology} & Describe the technologies that support oversight, such as monitoring dashboards, warning systems, or audit logs. Describe their:
  \begin{itemize}[noitemsep,topsep=2pt,partopsep=2pt,leftmargin=*]
   \item \textit{properties}: traceability; auditability, transparency, and legally robust.
   \item \textit{functionalities}: immutable logging systems that capture, aggregate and interpret relevant events to determine compliance;  replay tools to view potential non-compliant incidents; dashboards to provide real-time views of status and potential gaps,  risk registers; compliance tracking; auditing packages. Certification platforms would be used to certify people, systems, and organisations.
   \item \textit{failure modes}: Delayed data logging and aggregated data access, inability to inspect non-compliant behaviour, inability to trace data across systems.
   \item \textit{system interfaces}: dashboards for logged and aggregated data;  telemetry data; process and workflow definition and monitoring; checklists; attestation tools; data exchange between systems.
  \item \textit{signals}: compliance reports; interventions in the systemic and realtime layers; incident reports.
  \end{itemize} \\

3. & \textbf{Oversight Personnel} & Describe the personnel responsible for oversight, including:
  \begin{itemize}[noitemsep,topsep=2pt,partopsep=2pt,leftmargin=*]
 \item \textit{composition}: Internal compliance officers, safety assurance personnel; audit teams; quality assurance teams. 
   \item \textit{abilities/competencies}: Aviation standards and safety expertise; regulatory expertise; ability to make judgements under uncertainty.
   \item \textit{authority}: Demand evidence; mandate corrective interventions; approve/suspend systems or individuals; mandate reviews; escalate issues; prepare audit materials.
   \item \textit{responsibility}: Monitoring and maintaining compliance against requirements; data and documentation maintenance; identifying compliance gaps; coordinating corrective actions; supporting external regulatory.
   \item \textit{intentions}: Demonstrate compliance; reduce safety and operational risk; cooperate with external regulators; ensure AI-driven ATC systems align with standards and regulations.
  \end{itemize} \\

4. & \textbf{Oversight Organisation} & Describe the organisational setting in which the oversight takes place, including:
  \begin{itemize}[noitemsep,topsep=2pt,partopsep=2pt,leftmargin=*]
   \item \textit{rules and procedural constraints}: Procedures for compliance, safety management, auditing, documentation, and escalation.
   \item \textit{governance structures}: Audit committees; compliance and safety functions; clear separation between compliance and operational roles.
   \item \textit{reporting mechanisms}: Compliance reports; safety reports; audit fundings; incident reports; dashboards.
   \item \textit{coordination}: Coordinate with air services and ATC organisations to maintain compliance; with external regulators to understand requirements  and demonstrate compliance; with systemic and realtime layer personnel to ensure requirements interpreted and implemented correctly.
   \item \textit{constraints}: Aviation commission requirements and other regulations; time and real-time pressures.
  \end{itemize} 
\end{longtable}

\subsection{Oversight Process}
The oversight process describes all relevant activities in the oversight task. For each of the \textbf{real-time}, \textbf{systemic}, and \textbf{compliance} layers, define:

\subsubsection{Real-time Oversight Process}

\begin{longtable}{l@{~}p{0.12\textwidth}p{0.8\textwidth}}

 1. & \textbf{Monitor} &
Specify the types of monitoring activities and their temporal frequency: \begin{itemize}[noitemsep,topsep=2pt,partopsep=2pt,leftmargin=*]
 \item \textit{signals}: Monitoring happens through multiple channels, for AI and other artifacts in the ATC control room:
    (a) direct observation of flight conflict alerts and trajectory predictions (local, case-by-case monitoring);
    (b) peripheral awareness of unusual information (unexpected weather alerts, real-time error reports for the system); and
    (c) comparing to physical instrumentation and artifacts (e.g., flight strips, ground staff's alerts) to identify discrepancies and potential AI system errors.
 \item \textit{interpretation}: Controllers are highly training in how to interpret signals, and what potential failure modes could occur.
 \item \textit{frequency}: monitoring within sectors is continuous and real-time.
\end{itemize}\\

 2. & \textbf{Assess Risk} &
Define the risk assessment process to be used by the oversight personnel: 
\begin{itemize}
[noitemsep,topsep=2pt,partopsep=2pt,leftmargin=*]
    \item Controllers use the evidence from monitoring to establish the severity of safety risks. Effectively assessing risks requires pattern recognition on the part of the overseer(s): observing when AI recommendations deviate from expected ATC functions, typical pilot decisions, or local airspace considerations. Controllers use information about false alerts, implausible predictions, or inappropriate resolutions from AI to capture a growing severity of safety risks. Assessment also includes the controllers confidence in their own judgment (compared to AI predictions), recent history of AI system reliability, and cognitive capabilities and workload at the time of judgment.
    \end{itemize}  \\

 3. & \textbf{Decide} &
  Define the decision-making process that the oversight personnel follow, including: 
\begin{itemize}[noitemsep,topsep=2pt,partopsep=2pt,leftmargin=*]
 \item \textit{decision protocols}: Decision options include: (a) accepting the AI recommendation as-is, (b) overriding the AI recommendation and using the controller's own judgment instead; (c) modifying the AI recommendation (e.g., adjusting the parameters of a specific solution); (d) seeking verification from other stakeholders (e.g., pilots, supervisors, ground staff); (e) if high risk assessment severity, temporarily disabling specific AI functions for some situations and reverting to manual procedures; and (f) depending on severity, escalating to on-call technical support or logging for offline evaluation. Decision authority may also be distributed, with controllers making immediate decisions for their sectors, planning controllers coordinating effective handovers and traffic flow, and supervisors making decisions about system (de)activation.
   \item \textit{documentation}: Air-traffic control systems log every action made by controllers, including voice communications between controllers, pilots, and supervisors.
\end{itemize} \\

4. & \textbf{Intervene} & Specify intervention types and their temporal frequency (real-time, scheduled, or conditional).\\
   &   & See `Decide' above.

\end{longtable}

\subsubsection{Systemic Oversight Oversight Process}
\begin{longtable}{lp{0.12\textwidth}p{0.8\textwidth}}

 1. & \textbf{Monitor} &
Specify the types of monitoring activities and their temporal frequency:
\begin{itemize}[noitemsep,topsep=2pt,partopsep=2pt,leftmargin=*]
 \item \textit{signals}: system performance monitoring, including accuracy rates, error patterns, response times, and any updates to these (global system monitoring), incident \& safety reports, changes in performance, system downtime (if any), feedback from trials;
 \item \textit{interpretation}: system designers and maintainers interpret these signals by determine severity of individual incidents and software failures, identifying trends over time in incidents, air traffic flow, and late arrivals;
 \item \textit{frequency}: conditional in the case of safety incidents, scheduled (e.g., weekly or monthly review) for monitoring trends.
\end{itemize}\\

 2. & \textbf{Assess Risk} &
Define the risk assessment process to be used by the oversight personnel: 
\begin{itemize} 
[noitemsep,topsep=2pt,partopsep=2pt,leftmargin=*]
    \item  Risk assessment at this level is about immediate safety issues that could result in serious harm, as well as the potential for future issues in safety and air traffic flow. Risk severity would be focused on  safety, yet balanced against air traffic flow.  Oversight personnel assess risks such as model drift, data limitations, unsuitable interface design, alert fatigue, experience of controllers, and controller training.
\end{itemize}\\

 3. & \textbf{Decide} &
Define the decision-making process that the oversight personnel follow, including: \begin{itemize}[noitemsep,topsep=2pt,partopsep=2pt,leftmargin=*]
 \item \textit{decision protocols}: systemic decisions would be guided by ATC regulations and governance processes at high level, and protocols would be defined by national/international air services, as well as individual organisations producing ATC software;
 \item \textit{documentation}: incident reports, investigation reports, audit reports, software and process design decisions, software verification results, usability test results, and change logs.
\end{itemize}\\

 4. & \textbf{Intervene} &
Specify intervention types and their temporal frequency (real-time, scheduled, or conditional): In the case of serious safety incidents, interventions can be real-time; e.g. switch to the backup system. Most others are conditional, based on reporting, decisions, software upgrades, etc., and some scheduled; e.g. monitoring of safety and air traffic flow over time.\\

\end{longtable}

\subsubsection{Compliance Oversight Process}
\begin{longtable}{lp{0.12\textwidth}p{0.8\textwidth}}

 1. & \textbf{Monitor} &
 Specify the types of monitoring activities and their temporal frequency:
 \begin{itemize}[noitemsep,topsep=2pt,partopsep=2pt,leftmargin=*]
 \item \textit{signals}: reports on the systemic layers (internal audits, etc), safety incident reports, compliance dashboard information such as control status; aggregated data logs; evidence completeness metrics.
 \item \textit{interpretation}: signals are interpreted against requirements such as standards, regulations, governance requirements, internatl control objectives, etc;
 \item \textit{frequency}: continuous or periodic for ongoing controls; scheduled routine audits; conditional in response to serious incidents, system changes, and audit findings.
\end{itemize}\\

 2. & \textbf{Assess Risk} &
Define the risk assessment process to be used by the oversight personnel: At this level, risk that the lower layers do not comply with the relevant regulations, standards, etc. at the organisational level, considering changes to processes and AI/software/\\

 3. & \textbf{Decide} &
Define the decision-making process that the oversight personnel follow, including:
\begin{itemize}[noitemsep,topsep=2pt,partopsep=2pt,leftmargin=*]
 \item \textit{decision protocols}: internal compliance and procedures define thresholds for intervention; some decisions  are governed by specific processes, often by regulation;
 \item \textit{documentation}: records of compliance assessment; incident reports; audit reports; mandates to lower levels; reports on enforcement; etc., will be recorded and in most cases are mandated by regulation.
\end{itemize}\\

 4. & \textbf{Intervene} &
Specify intervention types and their temporal frequency (real-time, scheduled, or conditional):  initiating corrective actions when required (conditional); updating controls or procedures (scheduled/conditions); mandating additional evidence (conditional/scheduled); system access/user restrictions (realtime, conditional); escalation of issues (conditional).

\end{longtable}

\subsection{Cross-layer linkages and evaluation}

\begin{longtable}{l@{~}p{0.12\textwidth}p{0.8\textwidth}}

1. & \textbf{Information flows} & Information flows regularly between controllers, and from controllers to  supervisors, during runtime operation. When incidents occur, supervisors communicate these to  design, management, safety, and legal teams within the air services organisation, who then, if necessary, report to regulators. Similarly, regulators provide information on risks and new regulations, and the design, management, safety and legal teams communicate to controllers and supervisors about changes in procedure, protocol, etc. \\

2. & \textbf{Feedback loops} & Incidents that affect safety and traffic flow are reported upwards, and design teams manage incident logs, then are discussed and may become new features with an AI and software system. In most organisations, changes to systems are carefully monitored, and often need to go through simulation testing for several weeks without an incident before being put into practice. \\

3. & \textbf{Evaluation metrics} & Safety incidents, including the number and severity are the primary measure, with late arrival of aircraft a secondary measure. Errors within the AI system that do not lead to safety incidents are also logged and measured. At the regulatory level, metrics include certification compliance rates, audit findings, incident rates per flight operation, and serious incident investigations involving AI system failures. \\

4. & \textbf{Dependencies and feasibility} & Infeasible oversight is almost unheard of in ATC, however, in serious situations, such as ATC systems being unusable, many organisations have basic backup systems that provide just enough functionality for all aircraft to land safely as soon as possible. At such times, oversight of controllers is more difficult, and different procedures are put into place. When these infeasibility conditions are identified (e.g., through workload studied, cognitive task analysis, or incident investigations), findings are communicated through multiple channels: air service safety personnel inform ATC controllers and ground staff to implement operational restrictions; inform AI developers and vendors to redesign the algorithm or interfaces; and inform regulators through safety reports that may lead to certification requirement changes. \\

\end{longtable}
\section{Automated Insulin Delivery}

This appendix outlines an example of the human oversight framework applied to the domain of automated insulin delivery (AID) in Type 1 diabetes (T1D) management --- a safety-critical medical application involving real-time glucose monitoring and algorithmic insulin titration.

\subsection{Task Layer}

The task layer consists primarily of people living with T1D (end-users) who manage their condition daily, supported by healthcare professionals (HCPs) who provide clinical oversight. The system functions through a loop of integrated technological tools including continuous glucose monitors (CGM), insulin pumps, and control algorithms.

\begin{longtable}{l@{~}p{0.12\textwidth}p{0.8\textwidth}}

1. & \textbf{Object-Level Task} & Describe the object-level task to be performed:
  \begin{itemize}[noitemsep,topsep=2pt,partopsep=2pt,leftmargin=*]
   \item \textit{primary goals}: To maintain blood glucose levels within a target range (Time-in-Range) to avoid acute hypo- and hyperglycemia, prevent long-term micro- and macrovascular complications, and support normal daily activities as well as exercise.
   \item \textit{risks}: Severe hypoglycemia (leading to loss of consciousness/seizures) and diabetic ketoacidosis (DKA); risks are exacerbated by inaccurate dosing, delayed trend detection, or hardware/software malfunctions.
   \item \textit{outcomes}: Short-term safety (avoidance of extremes), long-term metabolic health (in the form of stable HbA1c), and the successful conversion of complex biological regulation into a manageable daily therapy where the cognitive workload remains sustainable for the patient.
   \end{itemize} \\

2. & \textbf{AI system} & Describe the AI system that supports or automates parts of this task:
   \begin{itemize}[noitemsep,topsep=2pt,partopsep=2pt,leftmargin=*]
   \item \textit{core properties}: Hybrid closed-loop systems integrate glucose sensing and delivery using control algorithms (e.g., PID or Model Predictive Control) to adjust basal insulin and predict future trajectories.
   \item \textit{functionality}: The system automatically modulates basal insulin delivery and provides recommendations for bolus dosing based on real-time and predicted glucose levels.
   \item \textit{inputs and outputs}: Inputs consist of real-time CGM data, carbohydrate estimates, and exercise markers; Outputs consist of automated insulin delivery adjustments and specific bolus dosing recommendations.
   \item \textit{level of autonomy}: High autonomy for basal control (often functioning while the user sleeps), but relies on user-initiated inputs for meal-time boluses and correction doses.
   \item \textit{known limitations} or \textit{failure modes}: Failure modes include sensor inaccuracies (e.g., compression lows), communication drops (Bluetooth failure), incorrect trajectory predictions, or inappropriate recommendations that increase hypoglycemia risk. Automation may decrease user vigilance or mask physiological symptoms.
   \end{itemize} \\

3. & \textbf{Human(s)} & Identify the people directly interacting with, using, or affected by the AI system:
   \begin{itemize}[noitemsep,topsep=2pt,partopsep=2pt,leftmargin=*]
   \item \textit{abilities/competencies}: End-users with T1D who must interpret trends and symptoms; HCPs with clinical expertise in therapy optimization. In pediatric settings, caretakers without formal medical education may act as the primary operators.
   \item \textit{degree of discretion or authority}: Users have the final authority to accept or override dosing suggestions and can suspend automation at any time to revert to manual mode.
   \item \textit{responsibilities}: Ensuring safe device operation, providing sufficiently correct carbohydrate estimation, and responding to system alerts and alarms.
   \item \textit{intentions}: Optimizing glycemic outcomes while maintaining immediate physical safety and long-term health.
   \end{itemize} \\

4. & \textbf{Organisation} & Describe the organisational setting in which the task takes place:
   \begin{itemize}[noitemsep,topsep=2pt,partopsep=2pt,leftmargin=*]
   \item \textit{hierarchical structures}: Clinical care teams prescribe and configure the systems; manufacturers provide the technology, technical support, and initial training.
   \item \textit{relevant rules and procedural constraints}: Usage is heavily regulated by medical device authorities (e.g., FDA, EMA/MDR); clinical treatment protocols guide system settings.
   \end{itemize}\\

\end{longtable}

\subsection{Oversight Layer}

\subsubsection{Real-time layer}

\begin{longtable}{l@{~}p{0.12\textwidth}p{0.8\textwidth}}

1. & \textbf{Oversight Task} & Describe the oversight task to be performed: 
  \begin{itemize}[noitemsep,topsep=2pt,partopsep=2pt,leftmargin=*]
   \item \textit{purpose}: To prevent severe hypoglycemia and DKA by reviewing algorithm outputs and intervening when therapy appears unsafe or misaligned with the user's actual physiological state.
   \item \textit{goals}: Safety-critical monitoring of insulin adjustments and immediate response to glycemic excursions.
   \item \textit{outcomes}: Successful avoidance of acute medical emergencies and maintenance of time-in-range.
   \item \textit{risks}: Risks include alarm fatigue, automation bias (trusting the system over physical symptoms), and failure to detect mechanical issues like pump occlusions.
   \end{itemize}\\

2. & \textbf{Oversight Technology (Optional)} & Describe the technologies that support oversight:
   \begin{itemize}[noitemsep,topsep=2pt,partopsep=2pt,leftmargin=*]
   \item \textit{properties}: Integrated software interfaces on insulin pumps and mobile applications.
   \item \textit{functionalities}: Predictive alert systems for hypo/hyperglycemia, "Insulin on Board" (IOB) tracking, and real-time CGM trend visualization (e.g., trend arrows).
   \item \textit{failure modes}: Warning systems may fail due to sensor inaccuracies, signal loss, or depleted device batteries.
   \item \textit{system interfaces}: Displays on the pump or smartphone showing current glucose, predicted trends, and insulin delivery logs.
   \item \textit{signals}: Auditory alarms (with increasing volume), haptic vibrations, and color-coded visual indicators.
   \end{itemize}\\

3. & \textbf{Oversight Personnel} & Describe the personnel responsible for oversight:
   \begin{itemize}[noitemsep,topsep=2pt,partopsep=2pt,leftmargin=*]
   \item \textit{composition}: Primarily the end-user (patient) and occasionally caregivers/parents or school nurses.
   \item \textit{abilities/competencies}: Proficiency in interpreting CGM trends and correlating device data with physical symptoms (e.g., shakiness or sweating).
   \item \textit{authority}: Direct authority to override suggestions, adjust targets, or disable the automated insulin delivery entirely.
   \item \textit{responsibility}: Monitoring system integrity (e.g., validity of sensor data) and responding to alerts.
   \item \textit{intentions}: Ensuring safety while minimizing the cognitive burden of disease management.
   \end{itemize} \\

4. & \textbf{Oversight Organisation} & As above.\\

\end{longtable}

\subsubsection{Systemic layer}

\begin{longtable}{l@{~}p{0.12\textwidth}p{0.8\textwidth}}

1. & \textbf{Object-level Task} & Describe the systemic oversight task:
  \begin{itemize}[noitemsep,topsep=2pt,partopsep=2pt,leftmargin=*]
   \item \textit{purpose}: Design oversight for traceability and safety analysis. It involves visibility into intermediate algorithm states (e.g., constraint limits) to understand the logic behind specific adjustments.
   \item \textit{goals}: Identifying systematic errors (e.g., recurrent post-meal hyperglycemia) and refining model assumptions to better fit the user's insulin sensitivity.
   \item \textit{outcomes}: Optimized clinical parameters and refined software performance.
   \item \textit{risks}: Risk of incorrect parameter tuning leading to chronic over- or under-delivery of insulin.
  \end{itemize} \\

2. & \textbf{Oversight Technology} & Describe the technologies that support oversight:
   \begin{itemize}[noitemsep,topsep=2pt,partopsep=2pt,leftmargin=*]
    \item \textit{properties}: Secure cloud-based data aggregation platforms.
    \item \textit{functionalities}: Retrospective data dashboards (e.g., Glooko, Libreview), immutable logs for auditing, and clinical simulators for testing parameter changes.
    \item \textit{failure modes}: Incomplete data logs due to syncing errors or unrecorded contextual factors like exercise or illness.
    \item \textit{system interfaces}: Web portals providing Ambulatory Glucose Profile (AGP) reports.
    \item \textit{signals}: Reports highlighting periods of "Time Below Range" or excessive system-initiated suspensions.
   \end{itemize} \\

3. & \textbf{Oversight Personnel} & Describe the personnel responsible for oversight:
   \begin{itemize}[noitemsep,topsep=2pt,partopsep=2pt,leftmargin=*]
    \item \textit{composition}: Clinicians with advanced technology training, safety engineers at device manufacturers, and AI data scientists.
    \item \textit{abilities/competencies}: Expertise in retrospective data analysis, root-cause analysis of device failures, and translating clinical insights into algorithmic refinements.
    \item \textit{authority}: Clinicians adjust foundational therapy settings; manufacturers push firmware updates.
    \item \textit{responsibility}: Analyzing aggregated performance patterns to ensure long-term system safety and efficacy.
    \item \textit{intentions}: Improving the accuracy and reliability of the automated therapy.
   \end{itemize} \\

\end{longtable}

\subsection{Oversight Process}

\subsubsection{Real-time Oversight Process}

\begin{longtable}{l@{~}p{0.12\textwidth}p{0.8\textwidth}}

1. & \textbf{Monitor} & Specify the types of monitoring activities and their temporal frequency: 
 \begin{itemize}[noitemsep,topsep=2pt,partopsep=2pt,leftmargin=*]
 \item \textit{signals}: Direct observation of CGM trends (visual arrows), tracking insulin delivery logs (IOB), and monitoring for physical symptoms of dysglycemia.
 \item \textit{interpretation}: Comparing AI dosing recommendations against physical sensations and recent contextual factors (e.g., meal size, activity level).
 \item \textit{frequency}: Monitoring is continuous and real-time, occurring multiple times per hour via device displays and alerts.
 \end{itemize}\\

2. & \textbf{Assess Risk} & Define the risk assessment process to be used by the oversight personnel: 
\begin{itemize}[noitemsep,topsep=2pt,partopsep=2pt,leftmargin=*]
    \item Users use pattern recognition to establish the severity of safety risks, observing when AI recommendations deviate from expected therapy or physical state. Risks are assessed by correlating device alerts with physiological symptoms. Assessment also includes the user's confidence in the current sensor accuracy and their own cognitive capacity at the time of judgment. 
\end{itemize}\\

3. & \textbf{Decide} & Define the decision-making process that the oversight personnel follow: 
 \begin{itemize}[noitemsep,topsep=2pt,partopsep=2pt,leftmargin=*]
 \item \textit{decision protocols}: Decision options include: (a) Accepting the AI recommendation; (b) Overriding with a manual bolus; (c) Adjusting temporary targets; (d) Performing a fingerstick check; or (e) Disabling automation.
 \item \textit{documentation}: Interactions are documented automatically via the pump's event log and cloud-synced therapy data.
 \end{itemize} \\

4. & \textbf{Intervene} & Specify intervention types and their temporal frequency: 
\begin{itemize}[noitemsep,topsep=2pt,partopsep=2pt,leftmargin=*]
    \item Execution of decisions occurs in real-time. This involves changing dosing, suspending the pump, or confirming values manually. Frequency is conditional based on the occurrence of glycemic excursions or system alerts.
\end{itemize} \\

\end{longtable}

\subsubsection{Systemic Oversight Process}

\begin{longtable}{l@{~}p{0.12\textwidth}p{0.8\textwidth}}

1. & \textbf{Monitor} & Specify the types of monitoring activities and their temporal frequency: 
 \begin{itemize}[noitemsep,topsep=2pt,partopsep=2pt,leftmargin=*]
 \item \textit{signals}: Monitoring global system performance via metrics like Time-in-Range, GMI, and incident reports.
 \item \textit{interpretation}: System maintainers and clinicians interpret these signals to identify trends in failure or "model drift" over time.
 \item \textit{frequency}: Scheduled on a quarterly basis for routine clinical reviews, and conditional upon the reporting of a safety incident.
 \end{itemize}\\

2. & \textbf{Assess Risk} & Define the risk assessment process: 
\begin{itemize}[noitemsep,topsep=2pt,partopsep=2pt,leftmargin=*]
    \item Personnel assess risks such as model drift, unsuitable interface design leading to errors, and user alert fatigue. Assessment focuses on potential safety issues that could lead to severe hypoglycemia or DKA across the user population.
\end{itemize} \\

3. & \textbf{Decide} & Define the decision-making process: 
 \begin{itemize}[noitemsep,topsep=2pt,partopsep=2pt,leftmargin=*]
 \item \textit{decision protocols}: Decisions are guided by clinical protocols and software design standards.
 \item \textit{documentation}: Incident reports, audit logs, and documentation of all clinical settings changes.
 \end{itemize}\\

4. & \textbf{Intervene} & Specify intervention types and their temporal frequency: 
\begin{itemize}
    \item Interventions include clinical parameter updates, software firmware patches, and user retraining. Frequency is typically scheduled (visits) or conditional (patches).
\end{itemize} \\

\end{longtable}

\section{Diagnostic Decision-Making}
\label{appendix:diagnostics}

This appendix outlines an example of the human oversight framework applied to diagnostic decision-making in radiology, focusing on head CT interpretation supported by an AI-enabled diagnostic decision-support system (e.g., stroke and haemorrhage detection). 

\subsection{Task Layer}
The task layer represents the socio-technical system of the primary activity that involves the AI system. Human oversight focuses on this level’s functioning and its potential risks.
\begin{longtable}{p{0.1\textwidth}p{0.9\textwidth}}

 \textbf{Object-Level Task}& Describe the object-level task to be performed:
  \begin{itemize}[noitemsep,topsep=2pt,partopsep=2pt,leftmargin=*]
   \item \textit{primary goals}: Timely and accurate interpretation of head CT images to support clinical decisions (e.g., identifying intracranial haemorrhage, early ischaemic change, mass effect, hydrocephalus), appropriate escalation of critical findings, and production of a clinically useful radiology report.
   \item \textit{risks}: Missed findings (false negatives) or incorrect findings (false positives) leading to delayed or inappropriate treatment (e.g., delayed neurosurgical referral, inappropriate stroke pathway activation), communication failures for critical results, and downstream patient harm; risks are exacerbated by time pressure, high caseloads, interruptions, limited clinical information, and image artefacts.
   \item \textit{outcomes}: High diagnostic accuracy for clinically important findings, good patient outcomes, minimum time-to-report (especially for emergency head CT), and consistency and clarity of reporting (e.g., following structured reporting templates, such as RadReport by RSNA).
   \end{itemize}\\

 \textbf{AI System}& Describe the AI system that supports or automates parts of this task:
   \begin{itemize}[noitemsep,topsep=2pt,partopsep=2pt,leftmargin=*]
   \item \textit{core properties}: AI-enabled decision-support integrated into PACS/RIS or a vendor platform; typical components include (a) image-processing and pre-processing pipeline (e.g., slice normalisation, artefact handling), (b) one or more trained models for detection/triage (e.g., haemorrhage detection, large vessel occlusion proxy markers, early ischaemic change support), and (c) user interface elements (e.g., overlays, flags, worklist prioritisation).
   \item \textit{functionality}: Triage and prioritisation of suspected abnormal head CTs, detection and localisation of acute findings (e.g., intracranial haemorrhage subtypes, midline shift, mass effect), and presentation of decision-support outputs to guide review.
   \item \textit{inputs and outputs}: Inputs include DICOM head CT image series, acquisition parameters, and optionally limited clinical metadata; outputs include worklist flags (e.g., ``suspected ICH''), probability or confidence scores, saliency/heatmap overlays or bounding boxes or segmentations, structured prompts (e.g., ``consider mass effect''), and pre-worded patient report.
   \item \textit{level of autonomy}: The AI system operates autonomously in generating alerts and prioritisation recommendations, but is intended as decision support; the radiologist retains authority for interpretation, changing and signing of reports, and escalation.
   \item \textit{known limitations} or \textit{failure modes}: Underperformance in cases with motion, beam-hardening, post-operative change or or poorly acquired imaging; atypical presentations; confounding anatomy (e.g., calcifications); distribution shift across scanners/protocols/patient populations; miscalibrated confidence leading to overtrust or underuse; false positives contributing to alert fatigue and workflow disruption; false negatives creating unwarranted reassurance if users rely on AI triage.
   \end{itemize}\\

 \textbf{Human(s)} & Identify the people directly interacting with, using, or affected by the AI system, such as operators or end-users. Describe their:
                       \begin{itemize}[noitemsep,topsep=2pt,partopsep=2pt,leftmargin=*]
                       \item \textit{abilities/competencies}: Radiologists’ expertise in neuroanatomy, pathology recognition, and clinical reasoning under uncertainty; ability to integrate imaging with clinical context and priors; competence in using PACS/RIS and interpreting AI outputs (including their limits) through formal trainging and/or experience working with decision support systems.
                       \item \textit{degree of discretion or authority}: Radiologists have discretion to accept, ignore, or override AI outputs, determine report content, and initiate escalation (e.g., critical results communication, stroke team notification, neurosurgical referral); referrers act on reports.
                       \item \textit{responsibilities}: Responsibility for accurate diagnosis and appropriate communication of urgent findings; maintaining professional standards in reporting; using AI outputs appropriately as supportive information rather than definitive diagnosis.                       \end{itemize}\\

 \textbf{Organisation}& Describe the organisational setting in which the task takes place, including:
                       \begin{itemize}[noitemsep,topsep=2pt,partopsep=2pt,leftmargin=*]
                       \item \textit{hierarchical structures}: radiology teams organised by shift and subspecialty; consultants/attendings, registrars/residents, and reporting radiographers (where applicable); clinical leads and modality leads; interfaces with emergency medicine, stroke services, neurosurgery, and critical care; on-call structures for out-of-hours reporting (can include outsourced teleradiology providers, potentially using different AI systems) and second opinions.
                       \item \textit{relevant rules, processes and procedural constraints}: protocols for head CT acquisition and interpretation (e.g., stroke pathways), outsourcing policies, critical result communication policies, reporting standards and documentation requirements, service-level turnaround targets, information governance, and local policies regarding AI use.
                       \end{itemize}\\

\end{longtable}

 \subsection{Oversight Layer}
The purpose of the oversight task is to monitor the operation of the AI system in context, to detect risks or failures, and to intervene when necessary. Oversight at this level ensures that the AI system functions safely, fairly, and as intended. 

\subsubsection{Real-time Oversight Task}
\begin{longtable}{p{0.1\textwidth}p{0.9\textwidth}}

 \textbf{Oversight Task} &
Describe the oversight task to be performed:
\begin{itemize}[noitemsep,topsep=2pt,partopsep=2pt,leftmargin=*]
 \item \textit{purpose}: Continuously monitor AI-supported head CT interpretation during active clinical work to detect emerging risks and prevent patient harm through intervention before it occurs.
 \item \textit{goals}: Identify inaccurate, misleading, or incomplete AI outputs at the patient case level; maintain alignment between AI output and clinical judgement.
 \item \textit{outcomes}: safe diagnostic decisions for individual patients, effective correction or override of erroneous AI outputs; timely escalation of AI failure patterns.
 \item \textit{risks}: Automation bias, alert fatigue, inappropriate reliance on AI under time pressure, delayed escalation when AI and human judgement conflict, and ambiguity about when intervention is required.
\end{itemize}\\

 \textbf{Oversight Technology (Optional)} &
Describe the technologies that support oversight, such as monitoring dashboards, warning systems, or audit logs. Describe their: \begin{itemize}[noitemsep,topsep=2pt,partopsep=2pt,leftmargin=*]
 \item \textit{properties}: Real-time availability, seamless integration into existing IT infrastructure, low interaction cost (e.g., minimum amount of additional clicks to access relevant information).
 \item \textit{functionalities}: Explainability features (e.g., sliency maps, bounding boxes, confidence indicators) which should help to identify inaccurate AI outputs more easily.
 \item \textit{failure modes}: Misleading saliency indicators, poorly calibrated confidence scores, excessive false positives contributing to alert fatigue, technical failures (e.g., where imaging does not reach the AI server), and poorly acquired images that do not meet minimum quality standards.
 \item \textit{system interfaces}: Direct integration with PACS viewers and reporting software.
 \item \textit{signals}: See functionalities.
\end{itemize}\\

 \textbf{Oversight Personnel} &
Describe the personnel responsible for oversight, including: \begin{itemize}[noitemsep,topsep=2pt,partopsep=2pt,leftmargin=*]
 \item \textit{composition}: reporting radiologists.
 \item \textit{abilities/competencies}: expertise in head CT interpretation, understanding of AI technology (intended use and performance boundaries/limitations of specific tools), ability to recognise AI failure modes, and capacity to integrate AI outputs with clinical context under time pressure.
 \item \textit{authority}: authority to override, ignore, or correct AI outputs and to initiate immediate clinical escalation.
 \item \textit{responsibility}: continuous case-level monitoring and intervention (i.e., ignore, overrride, stop displaying AI outputs) to prevent patient-level harm.
\end{itemize}\\

 \textbf{Oversight Organisation} &
Describe the organisational setting in which the oversight takes place, including:\begin{itemize}[noitemsep,topsep=2pt,partopsep=2pt,leftmargin=*]
 \item \textit{rules and procedural constraints}: local/national reporting standards and communication pathways/policies.
 \item \textit{governance structures}: departmental clinical governance arrangements.
 \item \textit{reporting mechanisms}: immediate discrepancy reporting;
 \item \textit{coordination}: coordination emergency medicine, stroke services, and neurosurgery (primary), as well as with IT (secondary).
 \item \textit{constraints}: time pressure, interruptions, and high workload.
\end{itemize}\\

\end{longtable}

\subsubsection{Systemic Oversight Task}
\begin{longtable}{p{0.1\textwidth}p{0.9\textwidth}}

 \textbf{Oversight Task} &
Describe the oversight task to be performed: \begin{itemize}[noitemsep,topsep=2pt,partopsep=2pt,leftmargin=*]
 \item \textit{purpose}: monitor AI-supported head CT workflows across cases and over time to detect systemic risks that are not visible at the individual case level (e.g., systematic bias among specific patient groups).
 \item \textit{goals}: Identify recurring error patterns, performance drift, misalignment between AI behaviour and clinical practice / clinical guidelines, or workflow disruptions introduced by AI use (e.g., exessive time spent on (a) false alersts; (b) fixing technical problems, or (c) workarounds).
 \item \textit{outcomes}: System-level interventions such as recalibration, retraining, redesign of AI models or interfaces, and modification of task (e.g., screening order), human (e.g., advanced training), or organisational (e.g., reporting protocols) attributes to prevent future harm.
 \item \textit{risks}: Delayed recognition of systemic failure, normalisation of deviance, diffusion of responsibility, and reliance on incomplete or poorly contextualised performance metrics.
\end{itemize}\\

 \textbf{Oversight Technology (Optional)} &
Describe the technologies that support oversight, such as monitoring dashboards, warning systems, or audit logs. Describe their: \begin{itemize}[noitemsep,topsep=2pt,partopsep=2pt,leftmargin=*]
 \item \textit{properties}: Longitudinal, aggregate, and auditable;
 \item \textit{functionalities}: performance dashboards, audit logs of AI outputs and clinician responses, case replay tools, and incident aggregation.
 \item \textit{failure modes}: Incomplete logging, lack of clinical granularity, aggregation masking clinically meaningful errors.
 \item \textit{system interfaces}: Data pipelines connecting PACS/RIS, EHR, AI vendor platforms, and quality and safety systems.
 \item \textit{signals}: Changes in alert rates, clusters of discrepancies, recurrent incident reports/tickets, unusal alorithm processing time, and performance shifts (e.g., changes in turn-around-times), unusual interaction patterns following updates or protocol changes.
\end{itemize}\\

 \textbf{Oversight Personnel} &
Describe the personnel responsible for oversight, including: \begin{itemize}[noitemsep,topsep=2pt,partopsep=2pt,leftmargin=*]
 \item \textit{composition}: Senior radiologists, modality leads, clinical safety and quality teams, and system designers.
 \item \textit{abilities/competencies}: Clinical risk assessment, audit and incident investigation skills, and understanding of AI system behaviour and performance limitations.
 \item \textit{authority}: Authority to modify deployment scope, suspend or restrict AI use, and mandate system updates or retraining.
 \item \textit{responsibility}: Identifying systemic risks, coordinating system-level interventions, and ensuring organisational learning.
 \item \textit{intentions}: Preventing future harm and maintaining alignment between AI systems and clinical practice.
\end{itemize}\\

 \textbf{Oversight Organisation} &
Describe the organisational setting in which the oversight takes place, including: \begin{itemize}[noitemsep,topsep=2pt,partopsep=2pt,leftmargin=*]
 \item \textit{rules and procedural constraints}: audit schedules, incident investigation protocols, and change-management processes.
 \item \textit{governance structures}: AI oversight committees and clinical safety boards;
 \item \textit{reporting mechanisms}: formal audit reports and safety reviews.
 \item \textit{coordination}: coordination between radiology, IT, vendors, and organisational leadership.
 \item \textit{constraints}: limited resources, vendor dependency, and poor IT infrastructure.
\end{itemize}\\

\end{longtable}

\subsubsection{Compliance Oversight Task}
\begin{longtable}{p{0.12\textwidth}p{0.8\textwidth}}

 \textbf{Oversight Task} &
Describe the oversight task to be performed: \begin{itemize}[noitemsep,topsep=2pt,partopsep=2pt,leftmargin=*]
 \item \textit{purpose}: Monitor AI-supported head CT systems to enforce regulatory, legal, and contractual obligations.
 \item \textit{goals}: Ensure compliance with medical device regulation (e.g., FDA or EMA), professional standards, and contractual requirements for safe and appropriate AI use.
 \item \textit{outcomes}: Verified compliance, mandated corrective actions (e.g., retraining or withdrawal), and sustained public trust.
 \item \textit{risks}: Regulatory lag, insufficient clinical specificity in compliance criteria, and misalignment between regulatory requirements and operational realities.
\end{itemize}\\

 \textbf{Oversight Technology (Optional)} &
Describe the technologies that support oversight, such as monitoring dashboards, warning systems, or audit logs. Describe their: \begin{itemize}[noitemsep,topsep=2pt,partopsep=2pt,leftmargin=*]
 \item \textit{properties}: Standardised, auditable, and legally robust.
 \item \textit{functionalities}: Compliance dashboards, certification records, audit trails, and adverse event reporting systems.
 \item \textit{failure modes}: Delayed reporting, incomplete data, or lack of traceability across systems.
 \item \textit{system interfaces}: Interfaces between healthcare organisations, vendors, and regulatory bodies.
 \item \textit{signals}: Compliance reports, mandated notifications, patient claims/medical negligence claims, and adverse event alerts.
\end{itemize}\\

 \textbf{Oversight Personnel} &
Describe the personnel responsible for oversight, including: \begin{itemize}[noitemsep,topsep=2pt,partopsep=2pt,leftmargin=*]
 \item \textit{composition}: Regulators, compliance officers, and accreditation bodies.
 \item \textit{abilities/competencies}: Regulatory expertise, audit capability, and safety evaluation.
 \item \textit{authority}: Authority to mandate retraining, restrict deployment, impose sanctions, or withdraw certification.
 \item \textit{responsibility}: Enforcing compliance and responding to safety signals that exceed organisational control.
\end{itemize}\\

 \textbf{Oversight Organisation} &
\begin{itemize}[noitemsep,topsep=2pt,partopsep=2pt,leftmargin=*]
 \item \textit{rules and procedural constraints}: Medical device regulation, professional standards, and contractual obligations.
 \item \textit{governance structures}: National and international regulatory frameworks.
 \item \textit{reporting mechanisms}: Mandatory incident reporting and audits.
 \item \textit{coordination}: Coordination with healthcare providers and AI vendors.
 \item \textit{constraints}: Jurisdictional boundaries and regulatory update cycles.
\end{itemize}\\

\end{longtable}

\subsection{Oversight Process}
The oversight process describes all relevant activities in the oversight task. For each of the \textbf{real-time}, \textbf{systemic}, and \textbf{compliance} layers, define:

\subsubsection{Real-time Oversight Process}
\begin{longtable}{p{0.12\textwidth}p{0.8\textwidth}}

 \textbf{Monitor} &
Specify the types of monitoring activities and their temporal frequency: \begin{itemize}[noitemsep,topsep=2pt,partopsep=2pt,leftmargin=*]
 \item \textit{signals}: AI-generated case-level outputs (e.g., predictions, flags, saliency maps, bounding boxes, confidence scores), image features observed during review, discrepancies between AI output and radiologist judgement;
 \item \textit{interpretation}: radiologists interpret signals through expert pattern recognition, triangulating AI outputs with image evidence, patient information, clinical context, prior imaging, and known AI failure modes;
 \item \textit{frequency}: continuous and real-time during active clinical practice.
\end{itemize}\\

 \textbf{Assess Risk} &
Define the risk assessment process to be used by the oversight personnel: 
\begin{itemize}[noitemsep,topsep=2pt,partopsep=2pt,leftmargin=*]
    \item Risk assessment focuses on the likelihood and severity of immediate diagnostic harm. Radiologists assess whether AI outputs plausibly reflect true pathology, the clinical consequences of acting or not acting on AI suggestions, and their own confidence under current cognitive and workload constraints. Disagreement between AI and human judgement increases perceived risk and triggers heightened scrutiny or verification.
\end{itemize}\\

 \textbf{Decide} &
Define the decision-making process that the oversight personnel follow, including: \begin{itemize}[noitemsep,topsep=2pt,partopsep=2pt,leftmargin=*]
 \item \textit{decision protocols}: primarily rely on human judgement informed by AI as a supportive signal; predefined escalation rules apply for critical findings, but accepting, rejecting, or ignoring AI outputs remains discretionary;
 \item \textit{documentation}: clinically significant discrepancies, critical result communications, and escalation actions are documented through reporting systems and local discrepancy mechanisms where required.
\end{itemize}\\

 \textbf{Intervene} &
Specify intervention types and their temporal frequency (real-time, scheduled, or conditional): 
Interventions are immediate and case-specific. They include overriding or ignoring AI outputs, re-examining image regions highlighted by AI, adjusting the order or priority of case review, seeking a second opinion, and initiating urgent clinical escalation based on clinical judgement.\\

\end{longtable}

\subsubsection{Systemic Oversight Oversight Process}
\begin{longtable}{p{0.12\textwidth}p{0.8\textwidth}}

 \textbf{Monitor} &
Specify the types of monitoring activities and their temporal frequency:
\begin{itemize}[noitemsep,topsep=2pt,partopsep=2pt,leftmargin=*]
 \item \textit{signals}: aggregated AI performance metrics (e.g., alert rates, false-positive/false-negative patterns), clusters of clinician overrides or discrepancies, rates of IT tickes, time taken to process the AI output, incident and near-miss reports, changes in operational context (e.g., software updates, protocol changes), unusal algorithm processing times, performance changes (e.g., changes in turn-around time);
 \item \textit{interpretation}: senior clinicians and safety teams interpret signals by identifying trends, correlations, and deviations from expected performance, and by distinguishing random variation from systematic bias or drift;
 \item \textit{frequency}: scheduled (e.g., weekly or monthly review) and conditional, triggered by incidents, sudden performance shifts, or system updates.
\end{itemize}\\

 \textbf{Risk Assess} &
Define the risk assessment process to be used by the oversight personnel: \begin{itemize} [noitemsep,topsep=2pt,partopsep=2pt,leftmargin=*]
    \item Risk assessment at this layer focuses on the potential for repeated or widespread harm. Oversight personnel assess whether observed issues indicate model drift, data mismatch, interface-induced misuse, training data limitations, or organisational factors such as workload or insufficient training. Risk severity is evaluated in terms of scale, persistence, and clinical impact.
\end{itemize}\\

 \textbf{Decide} &
Define the decision-making process that the oversight personnel follow, including: \begin{itemize}[noitemsep,topsep=2pt,partopsep=2pt,leftmargin=*]
 \item \textit{decision protocols}: decisions are guided by predefined governance processes (e.g., safety review boards) and may include human judgement informed by technical evidence and clinical expertise;
 \item \textit{documentation}: findings, rationales, and decisions are formally documented through audit reports, incident analyses, and change-management records.
\end{itemize}\\

 \textbf{Intervene} &
Specify intervention types and their temporal frequency (real-time, scheduled, or conditional): 
\begin{itemize}[noitemsep,topsep=2pt,partopsep=2pt,leftmargin=*]
    \item Interventions are system-level and preventative. They include recalibrating or retraining models, redesigning interfaces or alerts, modifying deployment scope or use conditions, updating local protocols and training, and, where necessary, suspending AI use pending further investigation.
\end{itemize}\\

\end{longtable}

\subsubsection{Compliance Oversight Process}
\begin{longtable}{p{0.12\textwidth}p{0.8\textwidth}}

 \textbf{Monitor} &
Specify the types of monitoring activities and their temporal frequency:
\begin{itemize}[noitemsep,topsep=2pt,partopsep=2pt,leftmargin=*]
 \item \textit{signals}: Compliance reports, audit findings, adverse event notifications, evidence of deviation from approved use or certified performance, and outputs from real-time and systemic oversight layers.
 \item \textit{interpretation}: Regulators and compliance officers interpret signals against formal requirements, approved indications, and risk classifications, rather than local clinical preferences.
 \item \textit{frequency}: Scheduled (e.g., periodic audits) and conditional (triggered by serious incidents or regulatory notifications).
\end{itemize}\\

 \textbf{Risk Assess} &
Define the risk assessment process to be used by the oversight personnel: \begin{itemize}[noitemsep,topsep=2pt,partopsep=2pt,leftmargin=*]
    \item Risk assessment focuses on regulatory non-compliance and unacceptable safety risk at scale. This includes assessing whether failures exceed allowable risk thresholds, indicate breaches of approved use, or undermine certification requirements.
\end{itemize}
\\

 \textbf{Decide} &
Define the decision-making process that the oversight personnel follow, including:
\begin{itemize}[noitemsep,topsep=2pt,partopsep=2pt,leftmargin=*]
 \item \textit{decision protocols}: Decisions are governed by formal regulatory procedures and legal mandates rather than discretionary clinical judgement.
 \item \textit{documentation}: Compliance decisions, enforcement actions, and rationales are formally recorded through regulatory and contractual documentation.
\end{itemize}\\

 \textbf{Intervene} &
Specify intervention types and their temporal frequency (real-time, scheduled, or conditional): \begin{itemize}[noitemsep,topsep=2pt,partopsep=2pt,leftmargin=*]
    \item Interventions enforce compliance and may include mandating retraining or re-certification, restricting or withdrawing approval, imposing reporting or monitoring requirements, or requiring organisational changes to ensure ongoing compliance.
\end{itemize}
\\

\end{longtable}

\end{document}